\documentclass[usedcolumn,usenatbib,usegraphicx]{mn2e}
\usepackage{amssymb}
\usepackage[totalwidth=480pt, totalheight=680pt,letterpaper]{geometry}
\usepackage[breaklinks, ps2pdf,backref,colorlinks=true]{hyperref}

\title[Smear fitting]{Smear fitting: a new image deconvolution method for
  interferometric data}
\author[R. I. Reid]{Robert I. Reid$^{1,2}$\thanks{Email:
    rob.reid@nrc-cnrc.gc.ca}\\
  $^1$Department of Astronomy \& Astrophysics, University of Toronto, 60
  St.\ George St.,\ Toronto, ON, Canada, M5S 3H8\\
  $^2$Dominion Radio Astrophysical Observatory, Herzberg Institute of
  Astrophysics, National Research Council, P.O. Box 248,\\
  \phantom{$^2$}Penticton, BC, Canada, V2A 6J9.}
\pagerange{\pageref{firstpage}--\pageref{lastpage}}
\date{Accepted}
\pubyear{2006}

\newcommand{\plotone}[1]{\centering
  \includegraphics[width=0.95\columnwidth,clip]{#1}}
\newcommand{\plotonescaled}[2]{\centering
  \includegraphics[width=#2\columnwidth,clip]{#1}}
\newcommand{\plottwo}[2]{\centering
  \includegraphics[width=0.49\columnwidth,clip]{#1}
  \hfil \includegraphics[width=0.49\columnwidth,clip]{#2}}

\newcommand{\csq}{\mbox{$\chi^2$}}
\newcommand{\textsub}[1]{_\textit{\scriptsize #1}}

\begin{document}
\label{firstpage}
\maketitle

\begin{abstract}
  A new technique is presented for producing images from interferometric data.
  The method, ``smear fitting'', makes the constraints necessary for
  interferometric imaging double as a model, with uncertainties, of the sky
  brightness distribution.  It does this by modelling the sky with a set of
  functions and then convolving each component with its own elliptical gaussian
  to account for the uncertainty in its shape and location that arises from
  noise.  This yields much sharper resolution than CLEAN for significantly
  detected features, without sacrificing any sensitivity.  Using appropriate
  functional forms for the components provides both a scientifically
  interesting model and imaging constraints that tend to be better than those
  used by traditional deconvolution methods.  This allows it to avoid the most
  serious problems that limit the imaging quality of those methods.
  Comparisons of smear fitting to CLEAN and maximum entropy are given, using
  both real and simulated observations.  It is also shown that the famous
  Rayleigh criterion (resolution = wavelength / baseline) is inappropriate for
  interferometers as it does not consider the reliability of the measurements.
\end{abstract}

\begin{keywords}
 techniques: image processing -- techniques: interferometric
\end{keywords}

\section{Introduction}
Interferometers give us a much sharper view than filled aperture telescopes of
the same area, by sampling the sky's spatial frequencies with a set of
baselines given by the separations between each receiver.  Unfortunately the
distribution of samples (``visibilities'') is incomplete, so the Fourier
transform of the measured visibilities is \emph{not} the sky brightness
distribution.  The Fourier transform of the visibilities instead yields the
``dirty map'', which is the sky brightness distribution plus noise, convolved
with the dirty beam (the Fourier transform of the sampling pattern).  In other
terms, a simple Fourier transform produces an image with absolutely no power at
unsampled spatial frequencies.  As a result the dirty beam exhibits sidelobes
causing the fainter structure in the image to be buried under the diffraction
patterns of the brightest objects in the field.

Unfortunately the most straightforward way of deconvolving away the effect of
the dirty beam would divide by zero in the $uv$ plane (the Fourier transform of
the image plane) wherever no measurement was made, and practical methods must
instead attempt to fill unmeasured regions of the $uv$ plane with a more
reasonable estimate than zero.  Simply interpolating between the visibilities
does not work in general because aliasing of oscillations in the $uv$ plane can
erroneously move the emission toward the center of the image plane, and distort
its appearance.  Thus there is no unique prescription for extracting the
optimum estimate of the true sky brightness, and many methods, most notably
CLEAN \citep{bib:hogbom74} and maximum entropy deconvolution
(\citealt{bib:gullskilling}, \citealt{bib:cbb}), that vary in their properties
have been devised.  They are usually, if not strictly correctly, called
deconvolution methods.  They rely on expected or desired properties of the
images such as positivity, locality, or smoothness to constrain the images they
produce.  This paper presents a new method, called smear fitting, which uses as
simple a model as it can as its main constraint, and can optionally use
additional constraints such as positivity and/or confinement of the source to a
certain region.  It renders the measurements into images with better fidelity
and resolution, and fewer image artifacts, than traditional deconvolution
methods.

A great benefit of smear fitting is considerably improved resolution for
objects with peak brightness greater than 4.20 times the root mean square (rms)
noise (it does not change the resolution of fainter features) without a loss of
sensitivity from reweighting the data.  This feature is extremely important
since better resolution cannot be achieved by simply adding data, and the noise
that comes with it, from longer baselines or decreasing the weight of short
baselines, without increasing the root mean square (rms) errors in surface
brightness over the whole image.  The objects of interest in typical radio
astronomical observations span wide ranges of both brightness and size, and
smear fitting offers a way to optimally handle both.  Smear fitting also
produces a set of components modelling the source, and calculates the
uncertainties of the distribution of those components on the sky.  The
components often correspond to distinct physical features of the source(s),
making their parameters and associated uncertainties of immediate scientific
interest.

Smear fitting has been implemented as a modification (patch) to difmap
\citep{bib:difmap1997}.  The patch, known as smerf, is freely available at
http:$/\!$/www.drao-ofr.hia-iha.nrc-cnrc.gc.ca/$\!\sim$rreid/smerf/.

\section{Smear Fitting}
\label{sec:smf}

\subsection{Procedure}
\label{subsec:procedure}

\begin{figure}
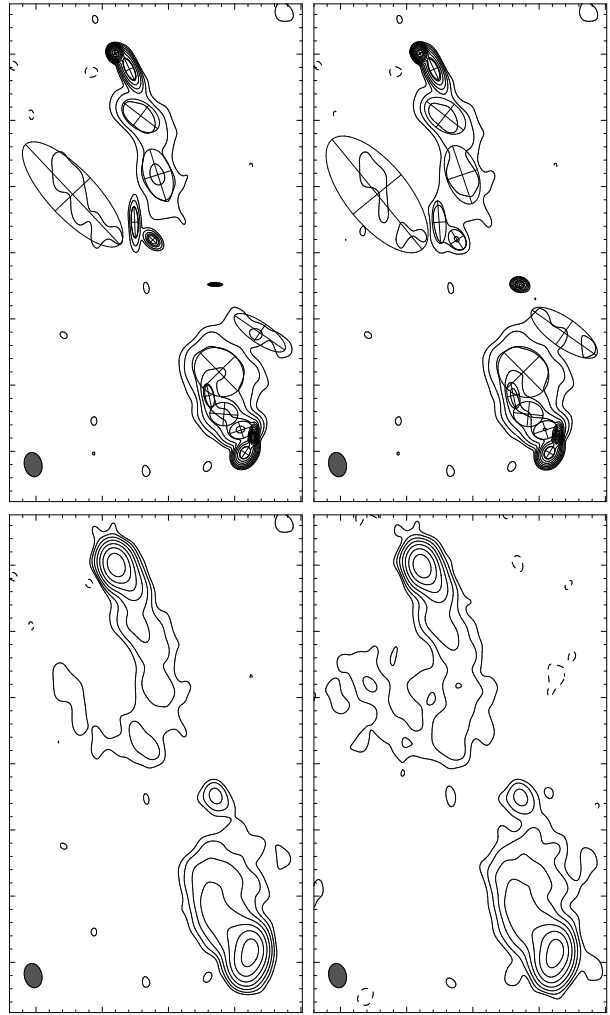

  \plottwo{0351-055unsm+c.ps}{0351-055sm+c.ps}\\
  \vspace{1mm}
  \plottwo{0351-055unsmcwcb.ps}{0351-055natcln.ps}
  \caption{Before and after smearing compared to CLEAN for a naturally weighted
  VLA snapshot observation of J0354--052 \citep{1999ApJS..124..285R}.  Top
    left: unsmeared image.  Top right: smeared image.  Note that smearing
    fattens the knife-edges, and that less significant components (shown as
    ellipses with FWHM major and minor axes) are smeared more.  Bottom left:
    unsmeared image convolved with the CLEAN beam.  Bottom right: CLEAN image.
    The contours start at 0.125 mJy/arcsec$^2$ and are each separated by a
    factor of 2.  The solid gray ellipse in the lower left corner of each image
    is the FWHM CLEAN beam.}
  \label{fig:compbefaft}
\end{figure}

Smear fitting is a two step process.  In the first step a set of components,
usually elliptical gaussians, with total visibility function ${\sf
  V}_{\mbox{\it \scriptsize model}}(u,v;\vec{p})$ is fitted to the visibilities
${\sf V}_i$ by varying the parameters $\vec{p}$ to minimize \csq:
\begin{equation}
  \label{eq:csq}
  \csq = \sum_i \left| \frac{{\sf V}_i - {\sf V}_{\mbox{\it \scriptsize
  model}}(u_i, v_i; \vec{p})}{\sigma_i}\right|^2
\end{equation}
where $\sigma_i$ is the uncertainty of measurement ${\sf V}_i$.  In the second
step each component is broadened, or ``smeared'', until \csq\ is raised by the
number of degrees of freedom of the component's distribution on the sky.  The
visibility model and residuals are then Fourier transformed and summed together
to form an image.  An example of the effect of smearing is shown in
Figure~\ref{fig:compbefaft}.

The goal of this process is a final map that shows what viewers intuitively
expect when looking at an image with resolution known to be imperfect: the true
image convolved by the probability distribution of where the radiation
originates from.  The smeared map is ideally equivalent to the average of an
ensemble of maps produced from all possible realizations of the noise added to
the measured visibilities.  The first implementation of smear fitting used the
Monte Carlo method, but the current technique of broadening the components
until \csq\ is raised by a certain amount is far more efficient.

In the first step, producing an unsmeared model, it is usually impossible to
specify a model in its entirety and then simultaneously fit all of its
parameters.  This is because typical initial dirty maps are dominated by a
small number of bright objects that must be modeled and removed before the
underlying structure becomes apparent, exactly as with CLEAN\@.  Fortunately
components that do not appreciably overlap can be independently fitted, as can
neighboring features with proper downweighting of the short baselines
\citep{bib:mythesis}.  Therefore the model is built incrementally (either
manually or by running a script) with cycles of:
\begin{enumerate}
\item Adding to the model one or more elliptical gaussian(s) for the brightest
  peak(s),
\item specifying the set of parameters that should be fitted (which
  includes parameters from previously fitted components that will be affected
  by the new component(s)), then
\item fitting them to the visibilities by minimizing \csq.
\end{enumerate}

A gaussian component fit to an unresolved source can approach a $\delta$
function, na\"ively extrapolating power to spatial frequencies far beyond
those sampled by the measurements.  This makes it necessary to somehow smooth
the model so that it does not claim a higher resolution than the measurements
warrant.  Smear fitting accomplishes that with its second step, ``smearing''.
During smearing, each component of the model is broadened by minimizing the
sharpness function while constraining \csq\ to rise to
$\chi^2_\textit{{\scriptsize unsmeared}} + \Delta$ using a Lagrange multiplier.
The sharpness function is
\begin{equation}
  \label{eq:B}
  B = \frac{2\pi}{u\textsub{max}^2}\frac{\sum_c f_c^2 / A_c}{\sum_c f_c^2}
\end{equation}
where $f_c$ and $A_c$ respectively are the flux and effective area of a
component $c$.  The effective area of a gaussian is $2\pi r\sigma^2 = \pi
ra^2/(4\ln 2)$, where $r$ is its (minor/major) axial ratio and $\sigma$ and $a$
are its standard deviation and full width at half maximum (FWHM) along its
major axis.  Only relative changes in $B$ matter, so it is convienient to scale
it using $u\textsub{max}$, the maximum baseline length in wavelengths.  That
makes the contribution to $B > 1$ for components sharper than the spatial
frequency corresponding to the longest baseline.  Squaring the fluxes allows
$B$ to be used with negative components as well as positive ones.

The form of $B$ came from devising a function to measure the squared amplitude
of the model visibilities integrated over the \emph{entire} $uv$ plane relative
to the squared total flux, with a modification to ignore component overlaps.
The larger $B$ is, the more of the model is in the part of the $uv$ plane
unsupported by measurements, so it is a quantity that should be minimized by
reliable models.  Each component's flux is held fixed during smearing to
prevent the sharpness decreasing by transferring flux between unrelated
components.

$\Delta$ is a positive number limiting the amount of smearing that is allowed.
The larger it is, the more smearing there will be, corresponding to a larger
confidence region of the model parameters.  Although model parameters are
often reported with confidence intervals of $\pm 2$, 3, or more standard
deviations, in the presentation of an image it is most natural to plot the
image ``smeared by one standard deviation'' in order to recreate the
probability distribution of the emission on the sky.  Another reason to only
use one standard deviation is that it is customary in radio interferometry to
publish images with whatever is not accounted for by the deconvolution (the
residuals) added to the result of the deconvolution.  Since smearing reduces
the amplitude of the long baseline model visibilities (broadening the image
plane distribution makes the $uv$ plane distribution more compact), it
effectively returns information from the model to the residuals.  The residual
map is of course dirty, so the final map will appear ``dirtier'' wherever the
effective smearing beam is comparable to or larger than the dirty beam.  This
is most likely not the desired effect, so it is important to distinguish
performing a reliable deconvolution of a set of data from displaying only the
information in that data that meets a certain level of reliability.  The
former is what smear fitting tries to do, but it can be adapted to the latter
task by displaying only the model in the final image, setting $\Delta$
appropriately, and making it clear that the resulting image is almost
certainly missing some real structure that did not ``make the cut''.

Finding $\Delta$ for a given confidence level is accomplished by considering
the Monte Carlo view of the process and requiring that the fraction of models
with \csq\ no greater than $\chi\textsub{orig}^2 + \Delta$, after being
perturbed by noise and refit, matches the confidence level.  Assuming gaussian
errors in the data, the model parameters will lie within $\pm 1$ standard
deviation of the best fit 68.3\% of the time.  Typically each component is an
elliptical gaussian smeared on its own, so the relevant degrees of freedom in
its distribution on the sky are its location (2 degrees), major axis, axial
ratio, and position angle, for a total of 5 degrees of freedom.  Solving
\begin{equation}
  \label{eq:68.3}
  0.683 = \int_0^\Delta P_5(\csq) d\csq
\end{equation}
with $P_5(\csq)$ being the probability distribution for \csq, yields $\Delta =
5.89$.  $\Delta$ for a 2 $\sigma$ confidence interval with 5 degrees
of freedom comes from solving
\begin{equation}
  \label{eq:95.4}
  0.954 = \int_0^\Delta P_5(\csq) d\csq
\end{equation}
and is 11.3.

\begin{figure}
  \plotonescaled{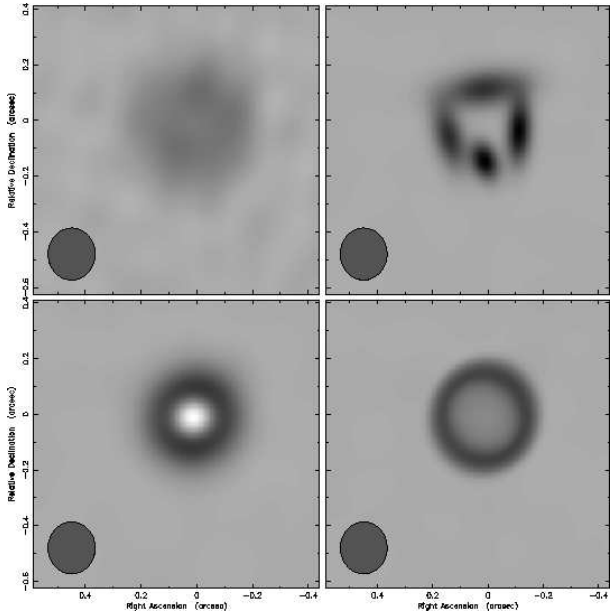}{1.0}
  \caption{A young planetary nebula, Vy2-2, as it appeared at 15 GHz in
    February 1982.  Top left: uniformly weighted CLEAN\@.  Top right: The
    result of automatic smear fitting with the {\tt modcons} script.  Bottom
    left: the result of fitting and then simultaneously smearing a positive and
    a negative gaussian.  Note the negative intensity in the center of the
    shell.  Bottom right: the result of fitting and then simultaneously
    smearing a positive and a negative optically thin ellipsoid (OTE).  Vy2-2
    is expected to be an optically thin shell, and it appears to be well
    modeled by the nested OTEs. The grayscale is a linear ramp from $-1.725$
    (white) to 3.510 (black) Jy/arcsec$^2$, and the flat gray ellipses are the
    FWHM beam extents.}
  \label{fig:vy22}
\end{figure}

To maintain the relation between convolving each feature with its uncertainty
and minimizing the sharpness while raising \csq\ by $\Delta$, it is usually
necessary to smear each component separately.  Otherwise a bright sharp
component would steal the portion of $\Delta$ that would normally go to a
fainter, broader, component, even if they were unrelated.  The only
justification for simultaneously smearing more than one component is when the
components are inseparable parts of a single feature.  In such (rare) cases
some customization is needed, either a modification of the sharpness function
to fairly distribute the smearing, or preferably replacement of the generic
elliptical gaussian components with a single component of a different type.
Currently implemented alternatives to elliptical gaussian components are
uniformly bright (i.e.\ optically thick) elliptical disks, optically thin
ellipsoids (Figure~\ref{fig:vy22}), and Sunyaev-Zel'dovich clusters \citep[p.\ 
351]{bib:pearsonnida}.

\newcommand{\rrmis}[1]{\mbox{{\it \scriptsize {#1}}}}

\subsection{The expected amount of smearing}
\label{sec:expsmearing}

A rule of thumb for the amount of smearing an isolated component will
receive can be generated using the following approximations:
\begin{enumerate}
\item The $uv$ sampling density is taken to be an elliptical gaussian,
  specifically the Fourier transform of the CLEAN beam.
\item Smearing is restricted to convolving the unsmeared component with a scaled
  version of the CLEAN beam with major axis $k$ times as large as the CLEAN
  beam's major axis, $a_b$.  
\item The component's unsmeared shape is assumed to have the same axial ratio
  and position angle as the CLEAN beam, so that it can be specified by the
  single parameter $\zeta \equiv a_c / a_b$, where $a_c$ is the component's
  major axis.
\end{enumerate}
Then the expected rise in \csq\ due to smearing is
\begin{eqnarray}
  \label{eq:simpexpdelcsq}
  \left< \Delta\csq\right> & = & \frac{2f_c^2 W}{1 +
  2\zeta^2}\frac{\xi^2}{1+3\xi+2\xi^2}, \\
  \mathrm{with}\ \xi & \equiv & \frac{k^2}{1 + 2\zeta^2}
\end{eqnarray}
and $W$ being the sum of the visibility weights.  

According to Equation~\ref{eq:simpexpdelcsq}, $\left< \Delta\csq\right>$
asymptotically approaches a maximum of $f_c^2 W/(1 + 2\zeta^2)^2$ for large
smearing beams.  This means that sufficiently diffuse components do not possess
enough signal to reach the target $\left< \Delta\csq\right>$ of $\Delta$, due
to a combination of insufficient flux and the fact that as components are
broadened they affect fewer visibilities.  Using the above approximations, the
limit at which smearing to a level of $\Delta$ cannot be done is
\begin{equation}
  \label{eq:smearlim}
  f_c^2W \simeq (1+2\zeta^2)\Delta
\end{equation}
For a component that is unresolved in the unsmeared map, this corresponds to a
surface brightness of $\sqrt{\Delta}$ times the theoretical r.m.s.\ noise of the
image.  If the component is significantly resolved, the surface brightness
threshold approaches $\sqrt{2\Delta}$ times the rms level.  In other words,
there is, unsurprisingly, a minimum surface brightness required for the
location and shape of a component to be determinable with any significance.
This holds true even when the above approximations are relaxed.

Thus a low surface brightness component can be completely smeared away without
sufficiently raising \csq.  In effect, it is completely returned to the
residual map, and should be removed to simplify the model.  If it contains a
significant amount of flux it should be CLEANed, since the dirty beam has no
total flux unless single dish measurements have been included.

\subsection{Instabilities in model construction and techniques to overcome them}
\label{sec:oversharp}

Most radio observations have objects in their field that are too compact to be
resolved with the given data, and as \csq\ is minimized, there is little to
stop the major and minor axes of the model components corresponding to those
objects from collapsing to zero.  Presumably all true sources have a nonzero
size, but the noise, since most of it originates in the $uv$ plane instead of
on the sky, does not necessarily obey the same rules as physical sources.  The
noise amplitude does not decrease with increasing baseline length (worse, it
gains weight since the visibilities become more isolated), so it is quite easy
for noise to collapse a compact component's shape (see the appendix).
When that happens the best thing to do is to fix the minor (and major, if
necessary) axes at tiny but nonzero values and to move on to other components.
This has no effect on the final image since the width of the smeared component
will be dominated by its smearing ``beam''.

The ``knife-edge'' case of a component with an extended major axis but
collapsed minor axis is particularly common since it does not require a
conspiracy of noise on both axes, and there are additional ways to produce it.
The original version of difmap featured an interesting shape parameterization
that unfortunately allowed the fit to enter a domain where the minor axis was
imaginary.  Once that happened the minor had to be clamped at zero and it
became very difficult for the fit to return to physical plausibility.  The
smerf patch to difmap features a better shape parameterization
(Equations~\ref{eq:alpha}~and~\ref{eq:gamma}) that allows the model fitting
routine to try any real value for the internal parameters $\alpha$ and $\gamma$
while keeping the full width half maximum of the major axis, $a_{\rrmis{maj}}$,
and axial ratio $r$ (ratio of the minor axis to the major axis), within their
physically allowable domains:
\begin{eqnarray}
  \label{eq:alpha}
  a_{\rrmis{maj}} & = & \frac{e^\alpha}{\pi u_{\rrmis{max}}}\\
  \label{eq:gamma}
  r & = & (\gamma^2 + 1)^{-1}
\end{eqnarray}
$u_{\rrmis{max}}$ is the longest baseline length.  

Also, knife-edges can appear where one model component has been used to fit
two unresolved features on the sky, so that \csq\ minimization joined the
features with a line.  smerf can automatically detect and remove such
knife-edge components, replace them with pairs of smaller components (if they
are sufficiently far apart), and reminimize \csq.

All of the above issues occur in the venerable model fitting stage of smear fitting, and
have fostered an impression in the community that fitting large sets of
elliptical gaussians to interferometer data is not feasible.  smerf includes
several features to improve the robustness of model fitting, but truly
unresolved sources are ubiquitous, and smerf's main method of dealing with them
is smearing.  Smearing broadens each component as much as possible given the
constraints of the data, smoothing $\delta$ functions and knife-edges as in
Figure~\ref{fig:compbefaft}, and at the same time adds to the image a
probability distribution for the locus of each component.  The smeared map
provides a very intuitive way to judge whether different components correspond
to significantly distinct features on the sky.

\section{Comparison To Established Methods}
\label{sec:compar}

\subsection{CLEAN}
\label{sec:clean}

The CLEAN process (\citealt{bib:hogbom74} and \citealt{bib:cbb}) is
similar to smear fitting except that it uses $\delta$ functions for the
components and a fixed gaussian for ``smearing'', which is called restoration
in the CLEAN nomenclature.  Most of the difference between the two methods in
the final image arises from the choice of smearing beam, but the philosophical
difference between the methods is in the way they interpolate between
visibilities in the $uv$ plane.  Smear fitting tries to make a gaussian go
through the error bars of each visibility, but CLEAN fits the flux in the
central visibilities and tapers off its gaussian according to the sampling
density (Figure~\ref{fig:uv-ampresponse}), even though the sky distribution is
unaffected by the sampling density.  In other words, CLEAN fails to fit the
measurements and tends to underestimate the achievable resolution.

\begin{figure}
  \plotone{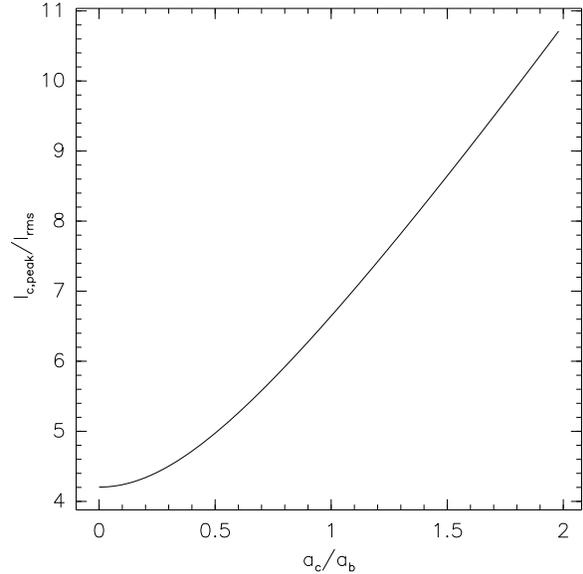}
  \caption{The surface brightness limit (relative to the theoretical root mean
    square noise) in the dirty map at which the smearing beam (one standard
    deviation) would be same size as the CLEAN beam for an elliptical gaussian,
    as a function of $\zeta$, the ratio of the unsmeared component major axis
    $a_c$ relative to the CLEAN beam major axis $a_b$.}
  \label{fig:snr}
\end{figure}

Smearing convolves faint components with a (potentially much) larger beam
than CLEAN\@.  Using Equation~\ref{eq:simpexpdelcsq} and its accompanying
approximations, the threshold at which the smearing beam becomes larger than
the CLEAN beam (Figure~\ref{fig:snr}) is
\begin{equation}
  \frac{I}{I\textsub{rms}} \simeq \left[
  (1+2\zeta^2)(3+5\zeta^2+2\zeta^4)\Delta \right]^{1/2} / (1 + \zeta^2)
  \label{eq:smearcln}
\end{equation}
For a component that is unresolved in the unsmeared map, this corresponds to a
surface brightness of $\sqrt{3\Delta}$ times the theoretical root mean square
noise of the dirty image.  For unresolved elliptical gaussians smeared by one
standard deviation, smearing is effectively equivalent to CLEAN at
approximately $\sqrt{3\Delta}\ (\simeq 4.20)$ times the rms surface brightness.
If the component is significantly resolved ($\zeta \rightarrow \infty$), the
threshold surface brightness ratio quickly approaches $2\zeta\sqrt{\Delta}\
(\simeq 4.85\zeta)$.  The smearing beam continues to enlarge below that
threshhold, becoming infinitely large as the surface brightness of the
unsmeared feature approaches zero.

Technically, this means that CLEAN is placing too much faith in the sharpness
and/or positional accuracy of the measurements for components below the
surface brightness limit in Equation~\ref{eq:smearcln}, but in imaging terms
smear fitting's behavior is very similar in the normal case, where the residuals
are added to the model for the final image.  When a component is severely
smeared, its model only accounts for the innermost visibilities, and leaves
the outer baselines in their original form.  In other words, the more a
component is smeared, the more the dirty beam shows through in the final image,
and the appearance of a radically smeared component in the final image is not
really broadened any more than it would be by CLEAN\@.  The size of the smearing
beam for each component can still be seen however, in a textual listing of the
smeared model.

\subsubsection{Total Flux Density and Source Counts}
\label{sec:totflux}

Although there is little cosmetic benefit to adding features below the limit of
Equation~\ref{eq:smearcln} to the model, it can be useful for both CLEAN and
smear fitting when the total flux density in the field is of interest, but no
zero baseline (i.e.\ single antenna) measurement is readily available.  In that
common case, the theoretical total flux density in the dirty beam, and thus the
residual map, is zero.  Practically, the positive part of the dirty beam is
concentrated in the central lobe and the negative part is spread over the rest
of the Fourier transformed region, so a sum of the residuals over a small area
will have nonzero flux density.  A more accurate estimate of the total flux
density can be obtained by modelling the trend of the short baseline
visibilities using a deconvolution method and including diffuse structure that
would not otherwise require deconvolution.  Both CLEAN and smear fitting can
include arbitrarily faint components in their models although systematic
effects can make those components unreliable.

As with CLEAN, a source count will of course be incomplete if model
construction is stopped early and only the component list (instead of the
image) is used for the count.  Similarly to flux estimation, either the model
construction stage of smear fitting can be continued to the required depth, or
a different deconvolution method may be used for the fainter emission.  More
interestingly, a single smear fitted component can correspond to many CLEAN
components, so a list of smear fitted components should provide a more accurate
raw count than a list of CLEAN components to the same depth.  More accuracy,
however, does not imply absolute accuracy.  Typically, but not always, smear
fitting allocates one component for each resolved feature, and the counter
still needs to decide which components should be coalesced into ``sources''.

\subsubsection{Visibility weighting}
\label{sec:weighting}

Since CLEAN depends so heavily on the sampling pattern, it is customary to
weight the outer visibilities more heavily than the inner ones, sacrificing
some sensitivity (especially to diffuse objects) for an improvement in
resolution \citep{bib:briggsphd}.  The basic idea of a common procedure,
called uniform weighting, is to weight each visibility by the reciprocal of
the number of visibilities within a certain distance of it.  Since long
baselines tend to be more isolated than short ones, uniform weighting -- or
anything more extreme, called superuniform weighting -- produces more compact
central lobes in the dirty beam, but the rms noise in the surface brightness
is degraded by the partial removal of short baseline data and the decrease in
beam area (the rms noise in the surface brightness is inversely proportional
to the product of the beam area and the square root of the number of
measurements).

Smear fitting does not require any reweighting in its final images, and
dynamically allocates the outer visibilities as much control as they deserve
based on their signal to noise ratios, so it achieves optimum sensitivity
without sacrificing any resolving power.

\subsubsection{Fixed vs.\ variable resolution}
\label{sec:rayleigh}

Smear fitting can, unlike CLEAN, adapt to scales smaller than the dirty beam
without losing sensitivity to diffuse structure (Figures~\ref{fig:sobclnvssf},
\ref{fig:012435smbsnap30}, and \ref{fig:012435cln30}).  In fact the most
striking difference between smear fitted images and CLEAN images is that smear
fitted images usually include one or more peaks that are much sharper than the
dirty beam.  It has been a long held precept that the smallest resolvable
angle, $\theta\textsub{min}$, is given by $\theta\textsub{min} = 1.2
\lambda/D$, where $D$ is the effective diameter of the telescope.  That
equation comes from the Rayleigh criterion for a circular aperture
\citep{bib:bornwolf7th}, which states that two equally bright point sources are
indistinguishable from a single source if their separation is less than the
distance between the diffraction pattern's central peak and its first minimum,
since at that distance most diffraction pattern convolved features appear to
meld together.  Deconvolution methods that reveal details finer than the
diffraction pattern are thus often called ``superresolving'', but since the
purpose of deconvolution is to remove the effect of the diffraction pattern,
the Rayleigh criterion is not the right one to use as a standard for
resolution.

In general the achievable resolving power of deconvolution is limited by how
well the effect of the diffraction pattern can be removed, which is determined
by both the size of the diffraction pattern's main lobe \emph{and} the signal to
noise ratios of the features to be resolved.  The blended beams of very close
features are difficult to disentangle unless the data have a high signal to
noise ratio, and conversely distant features are not reliably separated if one
or both are so faint that the probability distributions of their locations
significantly overlap.

\begin{figure}
  \plotone{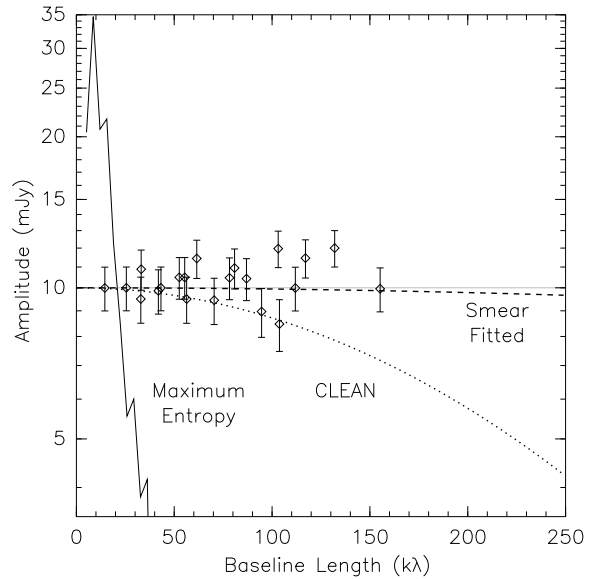}
  \caption{The response in the $uv$ plane of smear fitting,
  CLEAN, and maximum entropy to a simulated snapshot observation (error bars)
  of a 10 mJy point source (horizontal gray line) by a 7 antenna VLA
  subarray.  Maximum entropy
  did not perform well with this extremely sparse set of data, and unlike the
  other deconvolutions did not have the beneficial constraint of fitting a very
  simple model to this very simple source.  Also, its prior (a flat image) was
  completely inappropriate for a point source, so this should not be considered
  to be a typical case for maximum entropy deconvolution.}
  \label{fig:uv-ampresponse}
\end{figure}

\begin{figure}
  \plottwo{232327-3bnatcln.ps}{232327-3bunifcln.ps}\\
  \plottwo{232327-3bsm.ps}{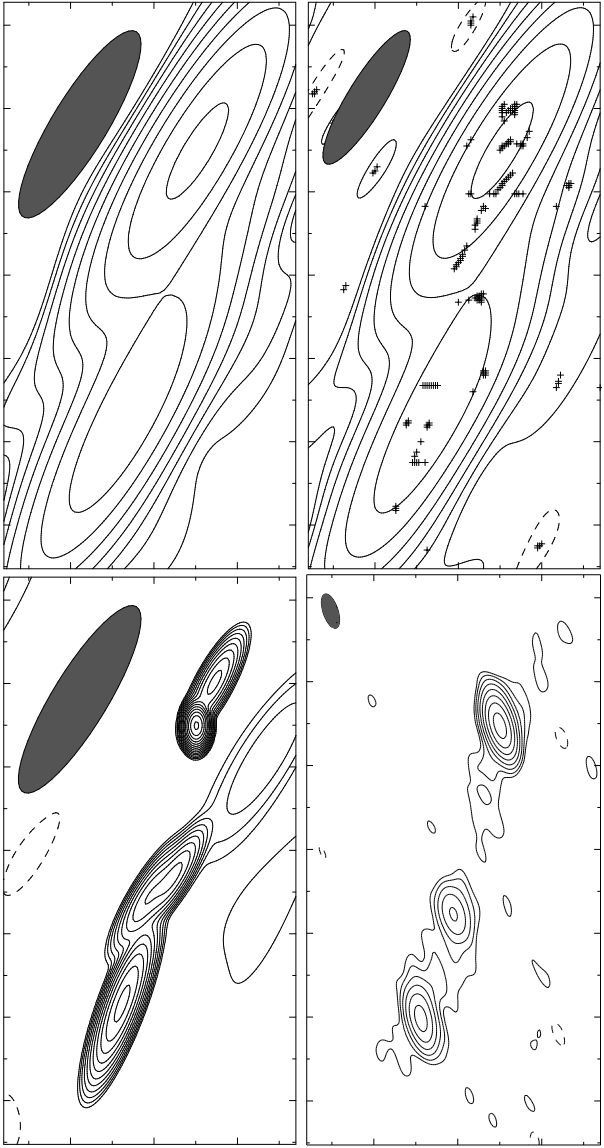}
  \caption{Top left: naturally weighted CLEAN image of a B configuration VLA 
    snapshot \citep{1999ApJS..124..285R}.  The lowest contour is at 31.25
    $\mu$Jy/arcsec$^2$.  Top right: uniformly weighted CLEAN image of the
    same snapshot.  The CLEAN components are shown as crosses, and hint that
    there are two jets and a core hidden by the CLEAN beams of the lobes.
    Bottom left: smear fitted image of the same data.  The contours start at
    31.25 $\mu$Jy/arcsec$^2$.  Bottom right: CLEAN image made with the addition
    of VLA A configuration data, confirming the reality of the structure
    suggested by the B configuration CLEAN components and made visible by the
    smear fitted map.  Note that some of the apparent discrepancy between the
    bottom images comes from the B array emphasizing emission from larger
    spatial scales than the A array.  The lowest contour is at 1
    mJy/arcsec$^2$.  In all of the images each contour is separated by a factor
    of 2 from its neighbor(s), and the gray disks show the FWHM of the dirty
    beams.}
  \label{fig:cleanvssm}
\end{figure}

As illustrated in Figure~\ref{fig:uv-ampresponse}, the resolution of the CLEAN
beam (and the Rayleigh criterion) is set by the average spatial frequency of
the measurements regardless of their values.  That is exactly what is needed
to emulate a nondeconvolved image from a filled aperture telescope of the same
size, since with filled aperture instruments there is one detector per
direction (for example, a pixel in a CCD) that simply receives the sum of all
the spatial frequencies sampled by the dish.  Interferometers have separate
measurements of the visibility function at the sampled spatial frequencies,
which the model fitting step of smear fitting uses to discern the trend of the
data.  The smearing step biases the model within the framework of its
parameters to be as broad as possible without deviating too far from the
data.  Figure~\ref{fig:cleanvssm} demonstrates that CLEAN is able to detect
sub-beamwidth structure which is obliterated by the restoring beam
in the final image.  Smear fitting is able to display the sub-beamwidth
structure, but without the often erroneous discreteness of the CLEAN component
distribution.

\subsubsection{Multiple scale CLEAN}
\label{sec:multiresclean}

\begin{figure}
  \plotone{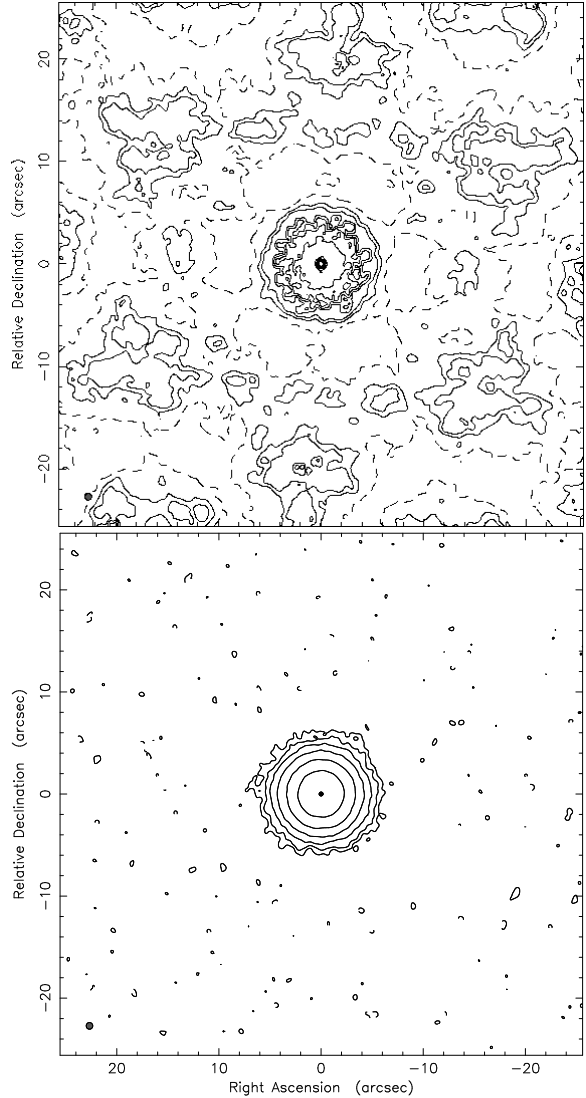}
  \plotone{sob2sm.ps}
  \caption{\label{fig:sobclnvssf}Deconvolved images of a simulated VLA snapshot
  of a 0.1 Jy, 0.1\arcsec\ FWHM, spike (central dot) on a 0.4 Jy,
  5\arcsec\ FWHM, circular
  gaussian.  Top: CLEAN\@.  The residuals are excessively large and reflect the
  structure of the dirty beam convolved with the broad feature.  Bottom: smear
  fitting.  The residuals are essentially uniformly distributed noise.  The
  model components for both images were automatically placed by a script
  without knowledge of the true source's structure.  The solid gray ellipse is
  the FWHM extent of the CLEAN beam.  The contours start at 0.25 mJy/arcsec$^2$
  and are each separated by a factor of 2.  Both methods used natural
  weighting.}
\end{figure}

\begin{figure}
  \centering
  \plotonescaled{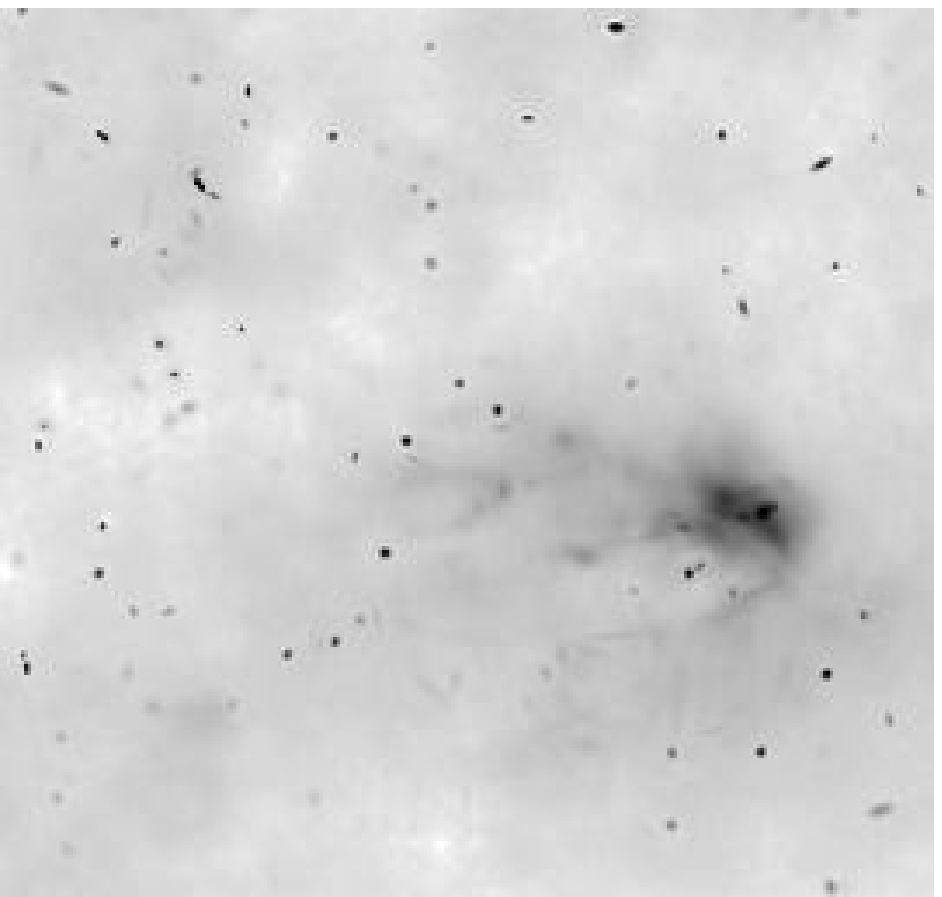}{0.8}\\
  \plotonescaled{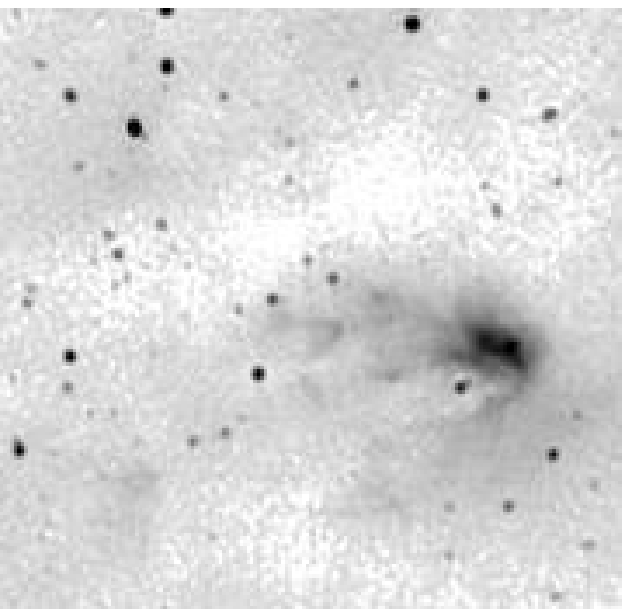}{0.8}\\
  \plotonescaled{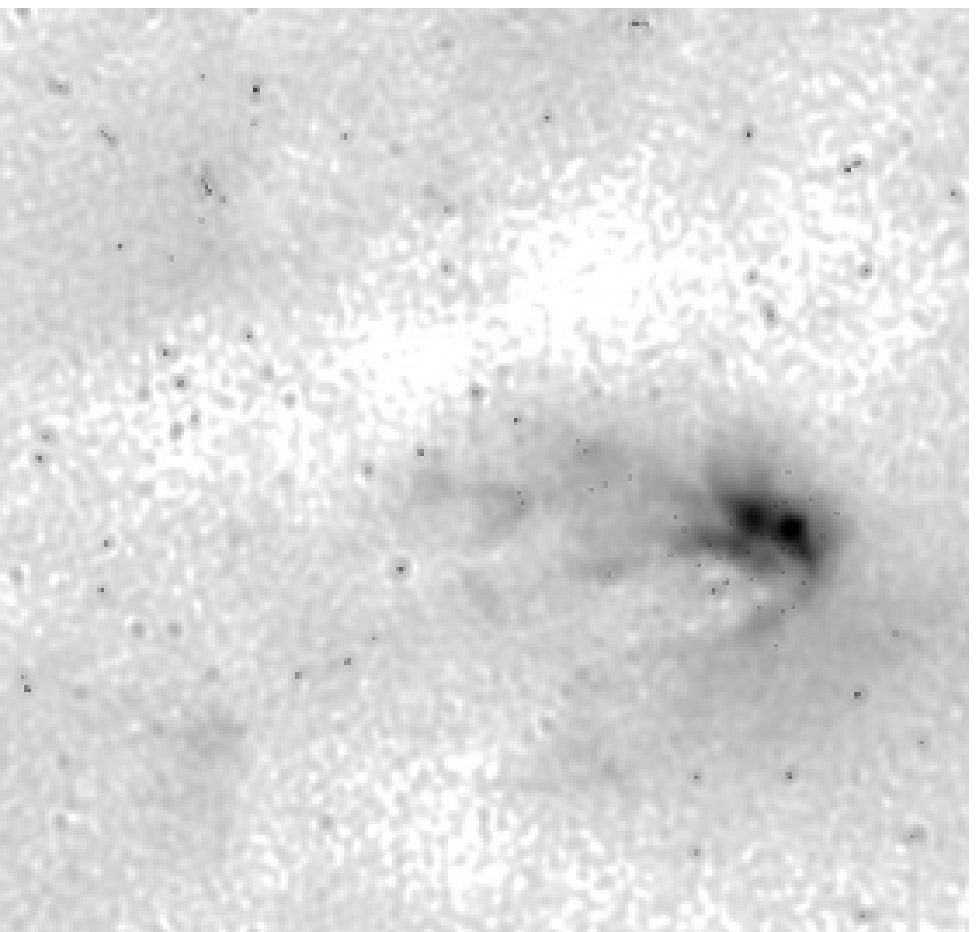}{0.8}
  \caption{Deconvolutions of a section of
  the Canadian Galactic Plane Survey \citep{bib:cgps2003} including many AGN
  and the less sharp galactic HII region Sharpless 155.  Top: smear fitted.  Middle: multiscale
  CLEANed and restored with a gaussian fit to the uniformly weighted dirty
  beam.  Bottom: multiscale CLEANed without convolution by a CLEAN beam.
  Convolution by the standard CLEAN beam obscures too much detail, but doing
  without it leaves the compact sources as $\delta$ functions embedded in the
  next larger gaussian used by multiscale CLEAN.  A
  square root function has been used for the grayscale intensity mapping.}
  \label{fig:msclean}
\end{figure}

``Multiresolution Cleaning'' \citep{mrc} is an extension to CLEAN that aims to
improve upon CLEAN's handling of extended emission (e.g.\ 
Figure~\ref{fig:sobclnvssf}) by using not just one CLEAN beam, but a set of
CLEAN beams of different scales.  In practice two variations, multi-scale CLEAN
\citep{bib:multiscaleclean} and adaptive scale pixel (ASP) decomposition
\citep{bhatnagar04:aspen1} work better and differ in how they use and set the
variously scaled beams.  For structures larger than the standard CLEAN beam
their behavior should fall somewhere between smear fitting and standard CLEAN if
natural weighting is used, with somewhat reduced flexibility because of the
finite number of choices.  Practitioners can however in principle preload the
algorithms with elongated gaussians if needed, and ASP dynamically updates its
set of scales.  Structures smaller than the standard CLEAN beam can be resolved
if sharp enough beams are used but such resolution (as in the bottom of
Figure~\ref{fig:msclean}) is not reliable since CLEAN and its derivatives do
not have a signal to noise based mechanism for testing resolution.  Much more
effort has been put into their ability to recover structures the size of the
standard CLEAN beam and larger.

\subsubsection{Self-calibration}
\label{sec:selfcal}

Smearing assumes that the measurement errors are normally distributed noise,
but the quality of many data sets is limited by calibration errors, which must
be corrected by selfcalibration \citet{selfcal1981}.  Selfcalibration assumes
(but is constrained in its action when the number of receivers is greater than
four) that the model includes most of the source and that the residuals are
mostly due to calibration errors.  Amplitude selfcalibration in particular can
be dangerous with classical CLEAN since traditional CLEAN picks the brightest
peaks first, and leaves behind any real structure that is fainter than the
calibration error induced artifacts around bright peaks.  This results in a
tendency of amplitude selfcalibration to distort the gains of central antennas
if applied carelessly.  Smear fitting and multiresolution CLEAN are both able
to fit broad features at an early stage, making it possible in most cases to
apply selfcalibration before calibration errors threaten to contaminate their
models and prevent convergence on the correct solution.

\subsubsection{Comparisons using known brightness distributions}

\begin{figure}
   \plotone{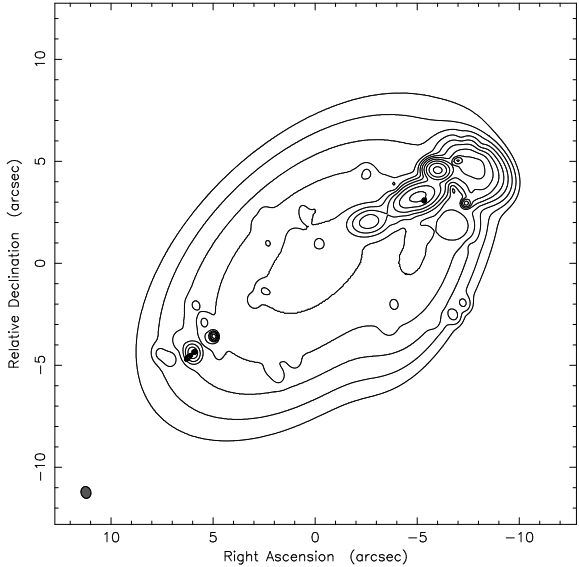}
   \caption{The reference image ``observed'' by the simulations, based on a
     smear fitted model of a 4.7 GHz VLA A and B array observation of the
     northern two--thirds of J012435-040105, with some faint (restored) CLEAN
     components, both positive and negative, included.  The contours start at
     20 $\mu$Jy/arcsec$^2$ and each is separated by a factor of 2 from its
     neighbor(s).} 
   \label{fig:012435ref}
\end{figure}

\begin{figure}
   \plotone{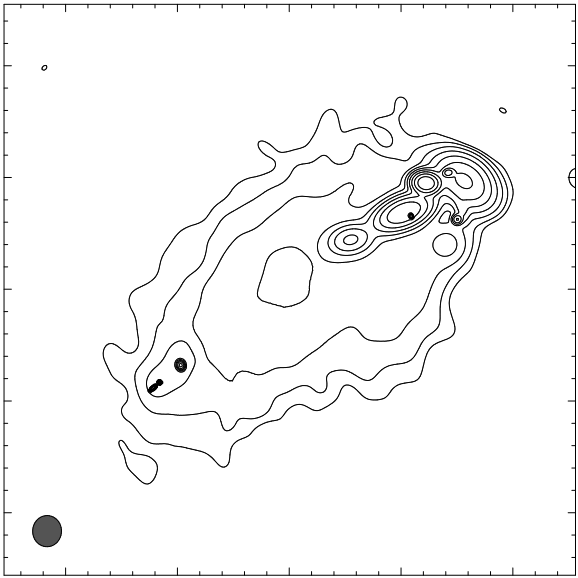}
   \caption{Image, using the true model, of a simulated 4.7 GHz VLA B array
     observation of Figure~\ref{fig:012435ref} placed at a declination of
   $+30^\circ$.  The contours start at 50 $\mu$Jy/arcsec$^2$ and each is
   separated by a factor of 2 from its neighbor(s).  The solid gray ellipse in
   the lower left corner shows the FWHM extent of the dirty beam.} 
   \label{fig:012435bsnaptruepnoise}
\end{figure}

\begin{figure}
   \plotone{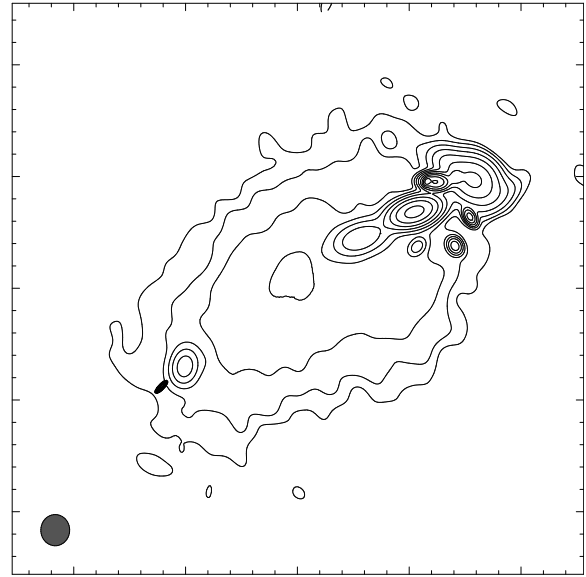}
   \caption{A smear fitted image made using the same data, contours, and beam as
     Figure~\ref{fig:012435bsnaptruepnoise}.} 
   \label{fig:012435smbsnap30}
\end{figure}

\begin{figure}
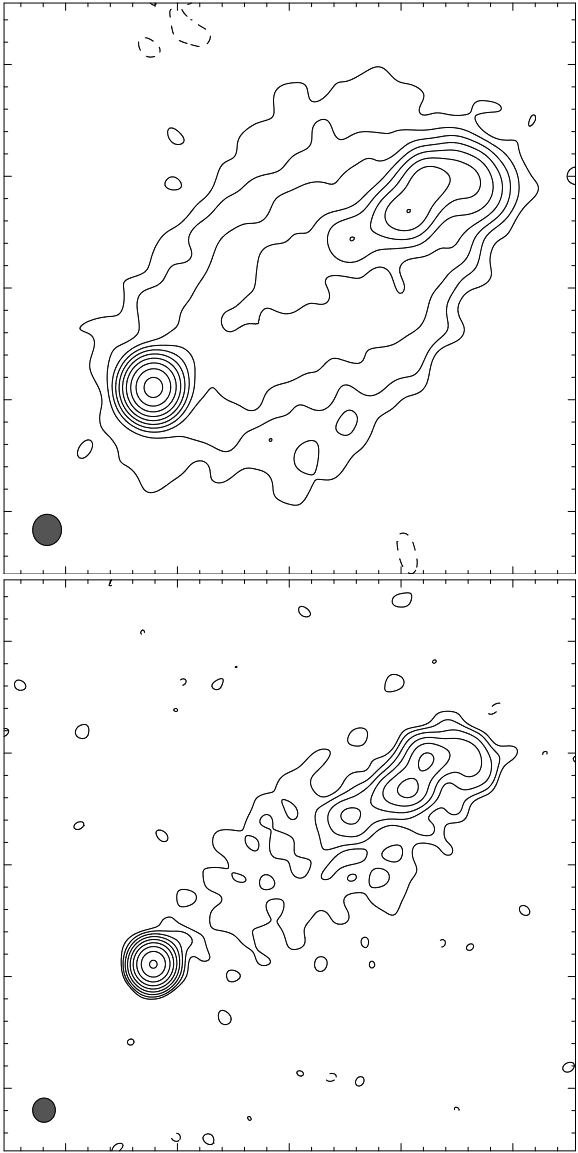

   \plotone{012435-0bsnap30natcln.ps}\\
   \plotone{012435-0bsnap30unifcln.ps}
   \caption{CLEAN deconvolutions of the same simulated observation used in
   Figure~\ref{fig:012435bsnaptruepnoise}.  Top: naturally weighted image with
   the same contours as Figure~\ref{fig:012435bsnaptruepnoise}.  Bottom:
   uniformly weighted image with contours starting at 0.2 mJy/arcsec$^2$, each
   separated from its neighbor(s) by a factor of 2.  The solid gray ellipses in
   the lower left corners shows the FWHM extent of the dirty beam.} 
   \label{fig:012435cln30}
\end{figure}

\begin{figure}
   \plotone{012435-0b8hsm.ps}\\
   \plotone{012435-0b8hunifcln.ps}
   \caption{Deconvolutions made from a simulated 8 hour VLA B array observation
   of Figure~\ref{fig:012435ref} placed at a declination of $+30^\circ$.  Top:
   smear fitted image with contours starting at 10 $\mu$Jy/arcsec$^2$.
   Bottom: uniformly weighted CLEAN image with contours starting at 50
   $\mu$Jy/arcsec$^2$.  Each contour is separated by a factor of 2 from its
   neighbor(s), and the solid gray ellipses in the lower left corner shows the
   FWHM extent of the dirty beams.} 
   \label{fig:012435b8h30}
\end{figure}

\begin{figure}
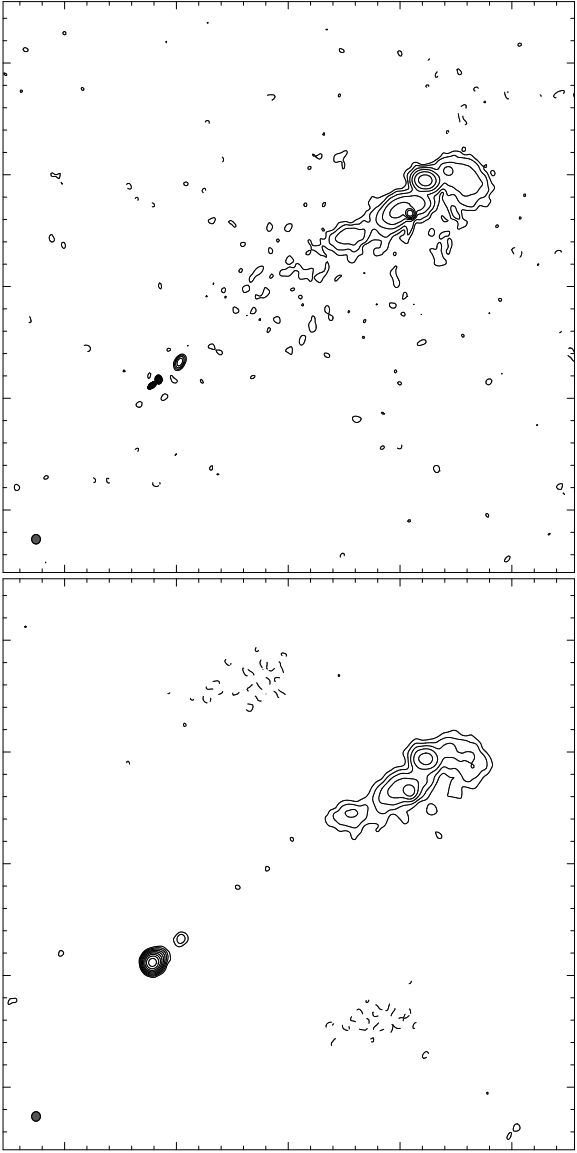

   \plotone{012435-0simasm.ps}\\
   \plotone{012435-0asnap30natcln.ps}
   \caption{Deconvolutions made from a simulated VLA A array snapshot of
     Figure~\ref{fig:012435ref} placed at a declination of  $+30^\circ$.  Top:
     smear fitted image with contours starting at 0.5 mJy/arcsec$^2$.  Bottom:
     naturally weighted CLEAN image with contours starting at 0.8
     mJy/arcsec$^2$.  Each contour is separated by a factor of 2 from its
   neighbor(s), and the solid gray ellipses in the lower left corner shows the
   FWHM extent of the dirty beams.} 
   \label{fig:012435sima}
\end{figure}

Figures~\ref{fig:012435ref} to \ref{fig:012435sima} present the responses of
CLEAN and smear fitting to a fairly complex source under a variety of
conditions.  In order that the true image could be known, and to dispense with
the effects of calibration errors, Miriad's \citep{bib:miriad1995} {\tt uvgen}
command was used to simulate VLA observations (with noise) of a given model.
To make the input model representative of a real observation, it includes the
northwest jet and core of a smear fitted model made with the combination of
several VLA A and B array snapshots of J012435--040105
\citep{1999ApJS..124..285R}.  A smear fitted model was used because it was the
best source of both very sharp and broad realistic structure available.  The
morphology of this source is also very challenging since it contains a sharp
feature of moderate flux density embedded in a broader stream, and the jet
bends over on itself into a crook.  The overall envelope of the jet was
enhanced by the addition of a broad elliptical gaussian.  To prevent the
model from being completely composed of smear fitting's stock in trade, several
dozen CLEAN components (restored with the A array beam), both positive and
negative, were randomly scattered around the jet.  Note that the CLEAN
component positions are completely random and not centered on pixels.

In the simulated B array observations (Figures~\ref{fig:012435bsnaptruepnoise}
to \ref{fig:012435b8h30}) the challenge was to resolve the three sharp peaks
near the core and as much of the jet morphology as possible.  Smear fitting did
much better, even when uniform weighting was used for CLEAN\@.  Neither of the
algorithms managed to resolve the two peaks closest to the core.  The smear
fitted image of the extended observation (Figure~\ref{fig:012435b8h30}) did
very well otherwise, while the CLEAN image was only able to improve its
sensitivity, not its resolution.  In theory the resolution of a smear fitted
image can improve without bound as data of the same average baseline length
is added, but in practice the improvement would eventually be limited by
calibration errors.

With the simulated VLA A array observation, Figure~\ref{fig:012435sima}, of
Figure~\ref{fig:012435ref}, smear fitting leaves the diffuse emission mainly to
the imagination of the viewer, although it could be argued that there is a
statistically significant enhancement of the density of small peaks in the
region of the diffuse emission.  Away from the emission region the residuals
are satisfyingly noise-like, but ironically the broad negative artifacts to the
north and south of the jet in the CLEAN image announce to the connoisseur the
possible existence of badly imaged diffuse emission.

\subsection{Maximum Entropy}
\label{sec:ME}

Smear fitting and CLEAN look for structure in the image plane and then impose
smoothness based on $uv$ plane considerations, but maximum entropy deconvolution
(\citealt{bib:cbb}, \citealt{CEMEM}), or ME, takes a different approach by
starting with a (usually) smooth default image (the prior) and trying to
maintain smoothness in the image plane as structure is imposed by the
visibilities.  Otherwise it has the same goal as smear fitting: to produce the
smoothest image possible within the constraints of the data.  In ME
deconvolution each pixel is considered to be a variable, but smear fitting can
be considered as a type of ME where the variables are a set of model
parameters.  This is not to say that smear fitting with an elliptical gaussian
for each pixel would be equivalent to traditional ME\@.  First, each
gaussian has 6 degrees of freedom, while a pixel only has one.  More
importantly, in traditional ME it is the distribution of pixel intensities
that is compared to the measurements, while smear fitting calculates \csq\ 
for the set of model parameters.

\begin{figure}
  \plotone{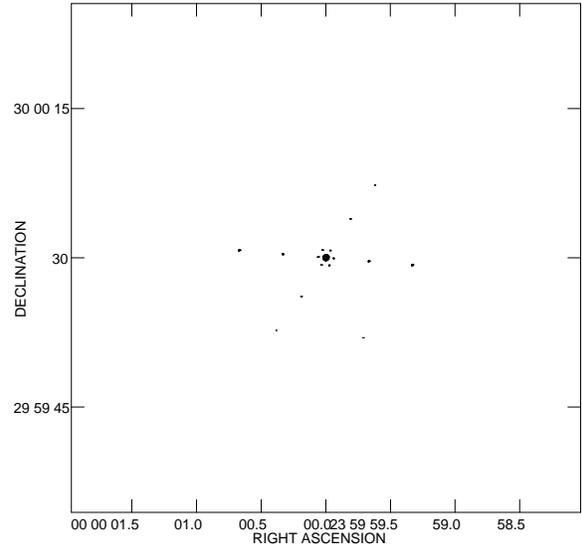}
  \caption{Ringing in a maximum entropy image, made using the AIPS 
    task VTESS, of the simulated data used in Figure~\ref{fig:sobclnvssf}.  The
    contours start at 40 mJy/arcsec$^2$ and are each separated by a factor of
    2.}
  \label{fig:sobmem}
\end{figure}

The choice of definition for the entropy function, ${\cal H}$, to be maximized, 
is controversial \citep{bib:nn1984} but the most popular one is
\begin{equation}
  \label{eq:IlnI}
  {\cal H} = -\sum_k I_k \ln\frac{I_k}{M_k e}
\end{equation}
where $I_k$ and $M_k$ respectively are the intensities of the $k$th image
and prior pixels.  Equation~\ref{eq:IlnI} can only be used for positive images,
but other functions which are between linear and quadratic, like
\begin{equation}
  \label{eq:lncoshI}
  {\cal H} = -\sum_k \ln\left( \cosh\frac{I_k - M_k}{\sigma} \right)
\end{equation}
where here $\sigma$ is the noise level in the image, also provide smoothing,
and have been used for polarization images.  All of these functions, however,
only consider the sum of functions of individual pixels.  This is hazardous
since the data, being in the $uv$ plane, are more directly connected with the
relationships between neighboring pixels than the values of individual pixels.
In particular, if the source has a sharp peak on top of a diffuse background,
the background dilutes the smoothing power that $\cal{H}$ would get from its
steepness at very low intensities, and the peak is left partly dirty
(Figure~\ref{fig:sobmem}).

Another criticism of traditional ME is that the default prior image, a flat
distribution, is actually a very \emph{unlikely} state for an interferometric
image (unless nothing has been observed!).  ME is thus biased toward putting
flux in pixels that should not have any, as in Figure~\ref{fig:uv-ampresponse}.
Smear fitting, like CLEAN, is more accepting of negative pixels, although both
often use positivity as a constraint when constructing their models.

Smear fitting and ME both produce final results that are biased away from the
best fit to the measurements, but they have their justifications.  ME has two,
the first being the importance of not claiming anything that is not absolutely
necessary, or in other words maximizing the entropy.  The second justification
is simply that ME produces better images with the bias than without it
(simulatable by making the uncertainties approach zero).  Smear fitting adds
the goal of trying to estimate the probability distribution of where the
measured light may have originated from.  Seen this way, it is not biased at
all, if the uncertainty determination is correct.

\begin{figure}
\newlength{\RRmaxencolwidth}
\setlength{\RRmaxencolwidth}{0.52\columnwidth}
\newlength{\RRsmcolwidth}
\setlength{\RRsmcolwidth}{0.47\columnwidth}
\begin{minipage}[t]{\RRmaxencolwidth}
  \begin{flushright}
    \includegraphics[height=0.99\RRmaxencolwidth, clip,
    angle=-90]{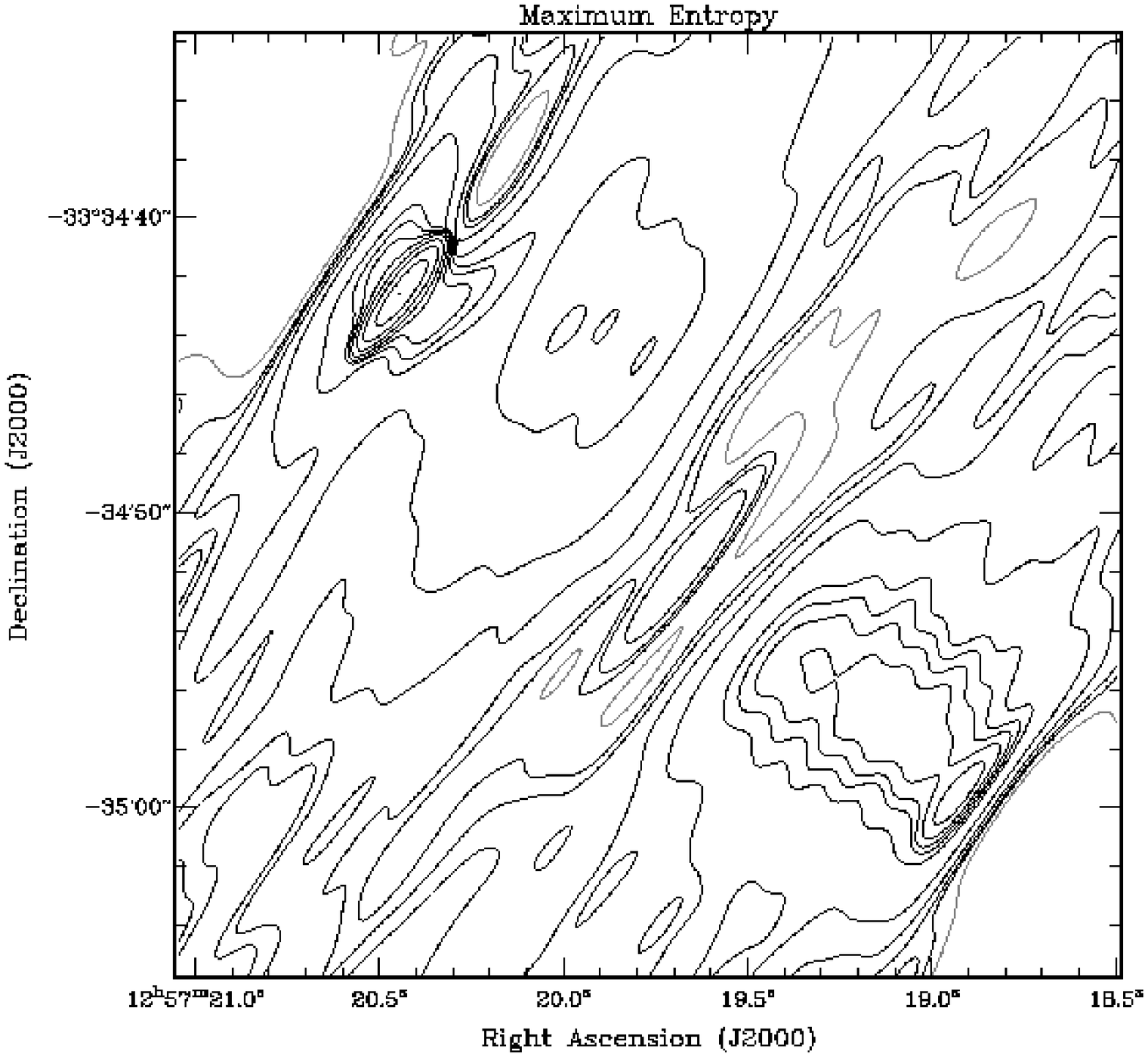}\\
    \includegraphics[width=0.894\RRmaxencolwidth, clip]{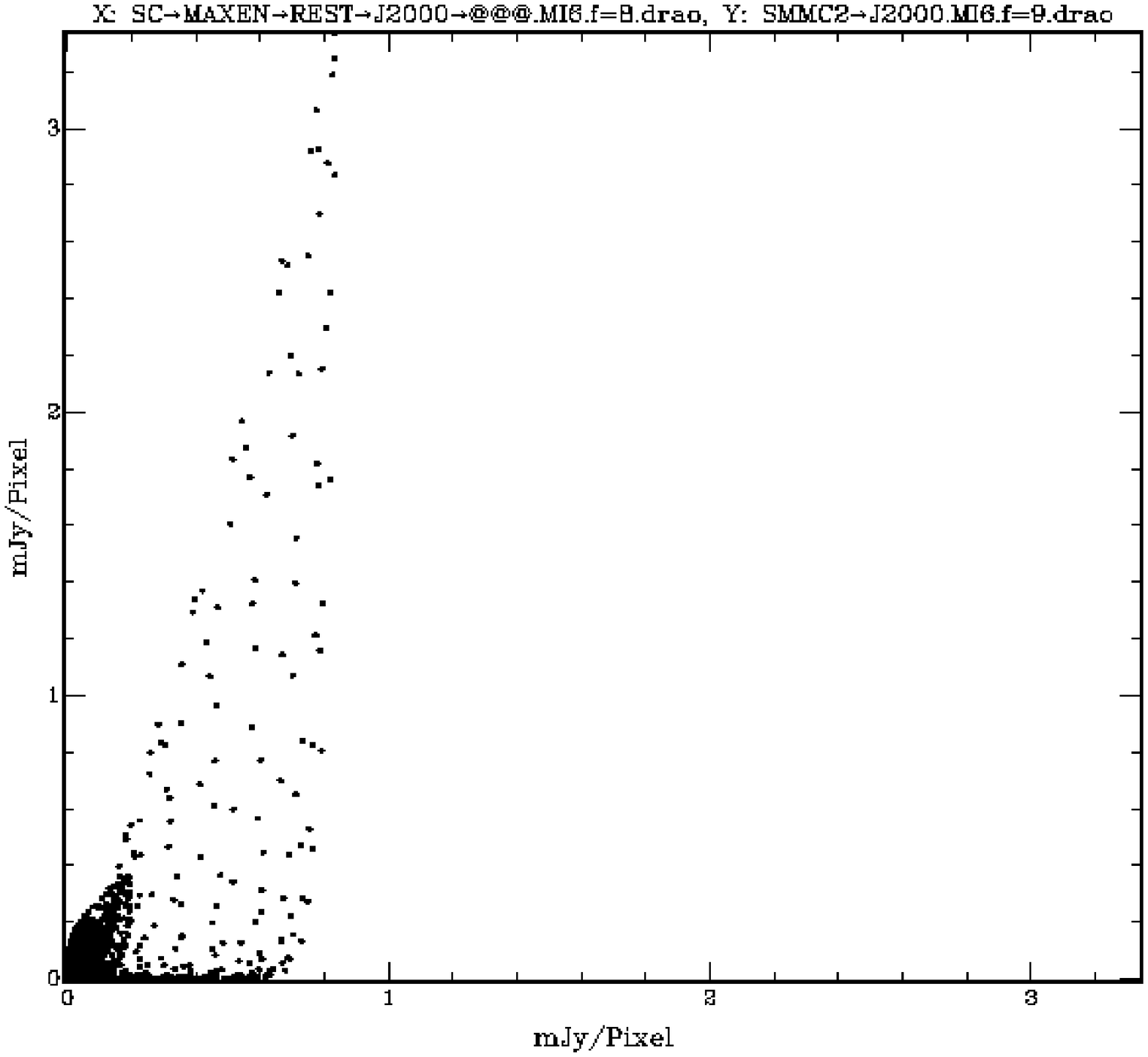}    
  \end{flushright}
\end{minipage}
\hfill
\begin{minipage}[t]{\RRsmcolwidth}
  \begin{flushright}
    \includegraphics[height=0.99\RRsmcolwidth,clip,angle=-90]{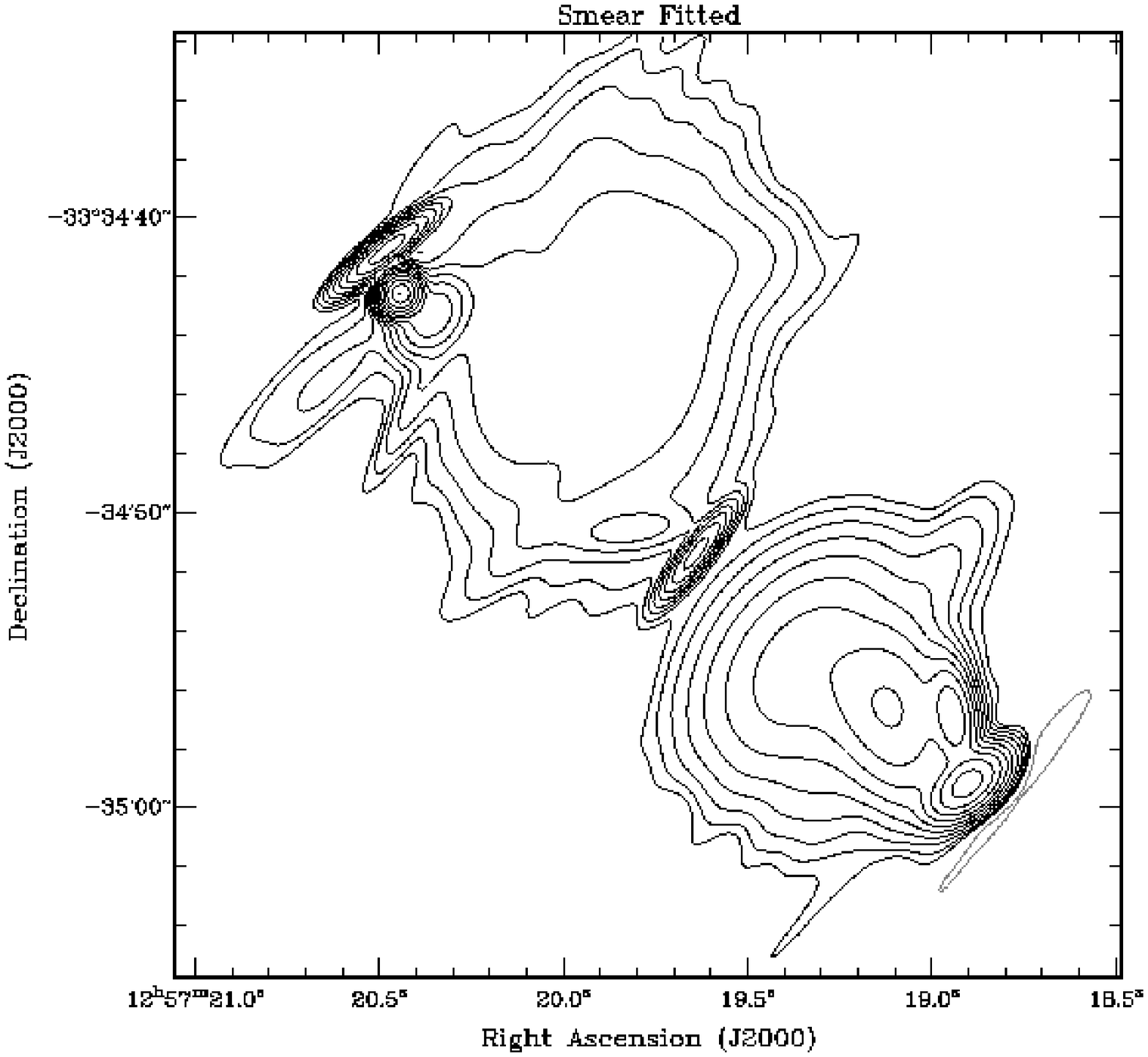}\\
    \vspace{1em}
    \includegraphics[height=0.99\RRsmcolwidth,clip,angle=-90]{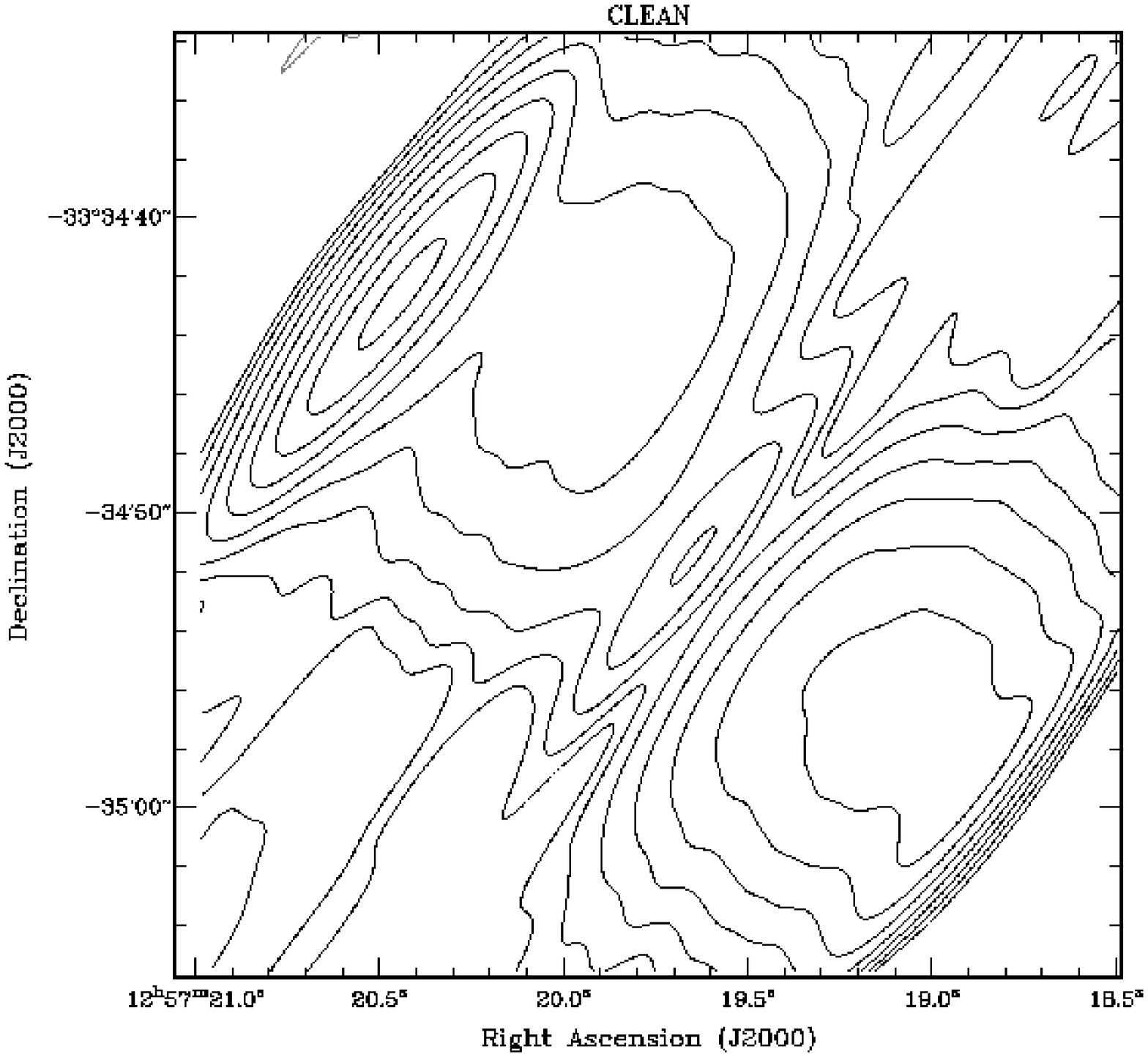}    
  \end{flushright}
\end{minipage}
\caption{Comparison of maximum entropy and CLEAN to smear fitting, for a
low elevation observation of J125720--333450 \citep{1999ApJS..124..285R}.
  The contours start at $\pm$ 0.032 mJy per square arcsecond, and increase by a
  factor of two between successive contours.  Each map used the same dataset,
  self--calibrated using smear fitting.  The bottom left graph plots the
  intensity of each pixel in the smear fitted (vertical axis) image against
  that of the same pixel in the maximum entropy (horizontal axis) image.}
  \label{fig:mevssmss}
\end{figure}

Figure~\ref{fig:mevssmss} illustrates the similarity of maximum entropy to
smear fitting, although it may not look like it at first glance since the
largest and most distracting contours are the faintest and least reliable.  It
was not feasible to deconvolve down to the noise, or even the rms level of the
smear fitting residuals, with maximum entropy, with any of the large sample of
VLA snapshots of jets in \citet{1999ApJS..124..285R}.  That is partly due to
the hot spots that all jets have.  Normally they would be CLEANed away before
applying maximum entropy, but as the CLEAN image shows, convolving the three
brightest points with the CLEAN beam would have hidden much of the structure.
Also, it could be that pixel--by--pixel flexibility of maximum entropy leaves
it more vulnerable to the sampling sparseness of the snapshot survey, which
could effectively act as an additional apparent noise source.

Nevertheless, when the lowest contours are ignored, the ME and smear fitted
maps are quite similar, and both show an effective beam for bright emission
that is much smaller and less elongated than CLEAN's.  The lower left corner of
Figure~\ref{fig:mevssmss} quantitatively compares them by graphing the
intensity of each smear fitted pixel against the intensity of the corresponding
pixel in the ME map.  If the images were identical, the locus of points would
be a straight line with slope 1.  Obviously the range of maximum entropy pixel
values is only about a quarter of that in the smear fitted image.  This cannot
be fully accounted for by the maximum entropy image being ``dirtier'', since
that mainly affects the faintest pixels.  This may be due to maximum entropy
treating entropy as a global property of the image, while smear fitting treats
entropy (smearing, roughly) as a local property of each component.  Smear
fitting thus prevented from wrongly transferring flux between separated
features in order to lower the overall sharpness of the image.  Indeed, the
sharpest components tend to have the most signal and are smeared least.

\begin{figure}
  \centering
  \plotone{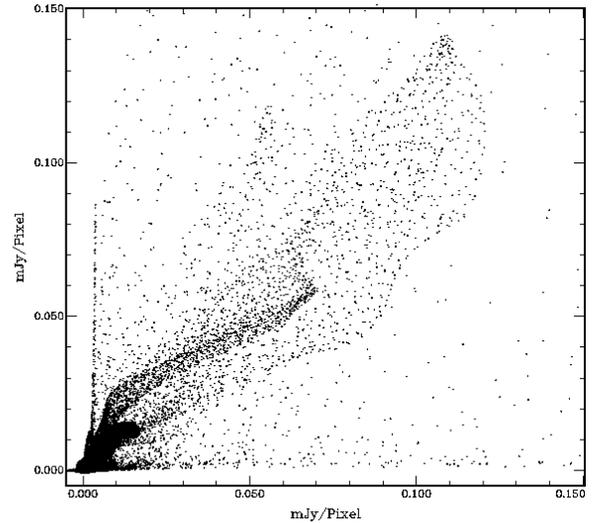}
  \caption{The fainter pixel intensities of the smear fitted (vertical axis) and ME (horizontal axis) images of Figure~\ref{fig:mevssmss}.}
  \label{fig:mevssmzoom}
\end{figure}

\begin{figure}
  \centering
  \plotone{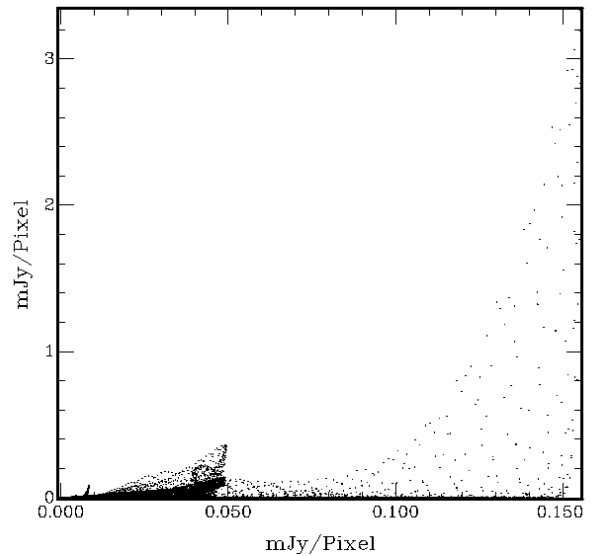}
  \caption{The pixel intensities of the smear fitted (vertical axis) and CLEAN
    (horizontal axis) images of Figure~\ref{fig:mevssmss}.  Note that the
    dynamic range of the CLEAN axis is much smaller than that of the smear
    fitted axis.}
  \label{fig:clnvssmint}
\end{figure}

Smear fitting and ME match more closely at medium intensities, as shown by
Figure~\ref{fig:mevssmzoom}.  CLEAN (Figure~\ref{fig:clnvssmint}), however,
does not approach a one-for-one match with smear fitting except at very low
intensities, where ME also has a short dense locus of points with overall slope
1 (the inner 15 $\mu$Jy of Figure~\ref{fig:mevssmzoom}).  Those pixels are in
the northwest lobe, where because it is broad and flat the pixel values are not
strongly affected by convolution with any beam the size of the CLEAN beam or
smaller.

\subsection{Pixel Based Models}

All existing deconvolution methods produce a rectangular array of pixel values
as their final image, but there is no fundamental reason why their internal
representation of the source should also be a rectangular array of pixel
values.  Although the desired output, an image capable of being displayed on
common computer hardware, is a set of pixels, the input measurements are an
irregular array of visibilities, and operations which need to consider
measurement uncertainty tend to be more practical when performed in the same
basis as the measurements.  One of the motivations of smear fitting was to
bring imaging conceptually closer to the measurements, and as a result the
models used in smear fitting have no explicit dependence on pixel size or
shape.

More prosaically, the locations of the component centers in smear fitting are
not quantized to land on pixel centers.  In many algorithms, including ME and
non-negative least squares (NNLS, \citet{bib:briggsphd}) deconvolution, the model is a
set of pixels, and in commonly available implementations of CLEAN the model
components are only placed on pixel centers (although in principle they do not
need to be).  Pixels poorly represent features that are centered on pixel edges,
which can degrade the usefulness of pixel based models for selfcalibration.
\citet{bib:hidrim} has also shown that a sharp feature centered on a pixel edge
cannot be represented by pixel centered CLEAN components without violating the
positivity constraint.

The Pixon deconvolution method~\citep{bib:pixon99} shares with smear fitting
the concept that the model (a preliminary image in Pixon's case) should be
smoothed as much as possible within the constraints of the data, and that the
amount of smoothing should be determined locally, not from a global prior as
with ME.  Instead of smear fitting's procedure of fitting a model to the
measurements by minimizing \csq\ and then smoothing that model, the Pixon
method convolves each pixel in a model image with the largest acceptable kernel
in a list of typically a dozen gaussians.  The Pixon technique thus uses a pair
of variables (flux and kernel size) for each pixel, while smear fitting eschews
pixels and attempts to model the source with the minimum number of parameters,
typically several dozen, required by the visibilities.

\csq\ minimization adjusts the model to maximize the probability of the data
given the model, so smear fitting could be broadly classified as a smoothed
maximum a posteriori probability (MAP) algorithm, while the Pixon method
results in a similarly smoothed NNLS image.  As a NNLS variant, the Pixon
method cannot represent sources with both positive and negative emission.
\citet{bhatnagar04:aspen1} also note that the Pixon method is not suited to
interferometry, since the algorithm relies on a compact point source response
function (i.e. a dirty beam which is zero outside a finite region), and noise
that is independent and additive in the image plane.

\subsection{Processing speed}
\label{sec:time}

Smear fitting tends to be somewhat slower than CLEAN, but not prohibitively so,
and most observations can be smear fitted using standard hardware in a
reasonable length of time.  Quantifying the speed of smear fitting relative
to CLEAN or Maximum Entropy can be done for a few examples, but extrapolating
from those examples to all observations is nearly impossible, because the speed
of smear fitting depends most strongly on the observation's ``complexity'',
which is itself difficult to measure.  The complexity increases with the number
of components, but also depends on how much coupling there is between the
components and how well the observation can be modeled with the given set of
basis functions.  To give some scale to ``not prohibitively'', using a 1.6 GHz
Athlon CPU, CLEAN took one minute to produce the naturally weighted image of
Figure~\ref{fig:012435cln30} while smear fitting took twenty minutes to
deconvolve the same simulated VLA snapshot (Figure~\ref{fig:012435smbsnap30}).
Neither algorithm was hampered by a lack of RAM or the need to self-calibrate.

To some extent the speed of other deconvolution methods is also affected by
source complexity, but to first order the processing time required by ME is
set by the number of pixels, while CLEAN's depends on the number of pixels
with significant brightness, being faster than Maximum Entropy for typical VLA
images with up to $10^6$ active pixels \citep{bib:cbb}.

Fitting a gaussian is more time consuming than placing a CLEAN component, but
one gaussian, or even better a specially selected function, in smear fitting
corresponds to many CLEAN components.  Typically CLEAN images have thousands
of components, while smear fitting only uses a few dozen, and the flux of each
CLEAN component is typically built up with 5 to 100 steps (a gain of 0.2 to
0.01).  Smear fitting can be the fastest method for simple but extended
sources and/or a small number ($\lesssim 50000$) of visibilities.  Note
however, that although the runtime of a single iteration of \csq\ minimization
is proportional to the number of visibilities, improving the dirty beam with
more $uv$ samples can make component specification more efficient and reduce
the number of  \csq\ minimization iterations needed.

The number of visibilities can be effectively lowered without drastically
degrading the dirty beam by binning them.  Normally binning the visibilities
should be avoided since it can create problems \citep{bib:briggsphd} and is not
required when smear fitting\footnote{Technically binning and gridding the
  visibilities is necessary
  for the FFT of the residuals in the final map, but any bright features should
  first be moved out of the residuals and into the model.}.  Fortunately large
data sets that are enough to make smear fitting annoyingly slow also tend to
have dense $uv$ coverage, eliminating many of the ambiguities that can impede
model construction, and making them good candidates for binning.  More
importantly, smear fitting does not use uniform weighting, which is responsible
for most of the imaging problems with binning.

\section{Discussion}
\label{sec:discussion}

Deconvolution by fitting simple models to the visibilities using \csq\ 
minimization is by no means new.  In fact it predates CLEAN
\citep{2003ASPC..300...17H} but garnered a reputation of being difficult and
unreliable.  The most serious problem with traditional model fitting, at least
for imaging, is that components corresponding to unresolved features can, and
probably will, collapse into Dirac $\delta$ distributions or knife-edges.
Smearing explicitly does away with that problem by convolving each component
with an elliptical gaussian set by the uncertainty of the component's shape and
location.  The other commonly heard complaint traditional model fitting is that
a model must be supplied before its parameters can be fit to the data, and in
the typical case of incomplete data it is impossible to be sure that the model
both has the right variables and started in the correct (i.e.\ global) local
minimum of \csq.  In other words, different astronomers can derive different
results from the same data because they started with different initial models
based on their subjective choices.  Of course, that situation is not unique to
imaging, but objectivity is still worth striving for.  Smear fitting removes
some of the subjectivity in the final image by smearing the statistically
insignificant details, but more pertinently its implementation as a patch to
difmap promotes automatic model construction.  Practitioners of smear fitting
should only rarely need to intervene in the model construction process, for
example by choosing to model a feature with an alternative functional form to
an elliptical gaussian, and in such cases should be able to support their
choice based on an improvement in \csq\ or positivity, or data from other
wavebands.

Smear fitting may appear to smear \emph{more} than CLEAN for low surface
brightness objects, but it must be remembered that whatever is smeared out of
the model is returned to the dirty residuals.  Usually the CLEAN beam size is
matched to the dirty beam, so smear fitting does not produce worse resolution
than CLEAN\footnote{The smear fitting analog of using an artifically small
  CLEAN beam would be to smear with $\Delta < 5.89$, i.e.\ convolve by
  $<$ 1 standard deviation.}.  Unfortunately the image is not the definitive
place to determine whether sub--beam features are \emph{extended}.  To properly
answer that question the reverse of smear fitting should be done; collapse the
component(s) down to a single Dirac $\delta$ distribution and check whether
\csq\ is raised above the minimum by at least twice the number of parameters
specifying the original shape.  The smerf patch does not provide a
command to automatically perform this check, but it is easy enough to do on a
case-by-case basis.  These properties of variable resolution also apply to ME
if it is not convolved with a CLEAN beam at the end\citep{bib:cbb}, although
the test for resolution, by minimizing the entropy, would in general not be as
useful since ME works on an entire image at a time instead of specific features.

\subsection{Sensitivity and weighting}
\label{sec:sensitivity}

Although smear fitting cannot \emph{deconvolve} low surface brightness features
any better than the other methods, its use is beneficial to being able to
\emph{detect} them.  Smear fitting uses the most sensitive weighting scheme,
natural weighting, while the other methods often downweight the inner
visibilities to produce a sharper and more gaussian dirty beam.  Robust
weighting \citep{bib:briggsphd} is a considerable improvement over uniform
weighting, but natural weighting still gives the greatest surface brightness
sensitivity.  All deconvolution methods can use natural weighting, but it
produces a large beam, and no fixed resolution method can distinguish faint
emission from bright peaks when they are within a beamwidth of each other.

Uniform or superuniform weighting hurts the sensitivity to sharp as well as
diffuse objects by effectively ignoring central visibilities even though they
measure the flux of sharp features just as well as the long baseline
visibilities (ignoring possible differences in antenna sensitivities).  Smear
fitting uses all visibilities for detection of flux, so that the position of a
visibility simply affects its leverage on the resolution.  

More subtly, smear fitting avoids the most common errors in fitting the
\emph{bright} features that typically limit the dynamic range of images
deconvolved using other methods.  The first two of these errors are due to
premature pixelization in order to use the Fast Fourier Transform (FFT).  The
FFT requires the visibilities to be placed in rectangular bins, and outputs the
result in rectangular pixels.  This introduces firstly a quantization error to
the positions of the visibilities. Secondly, if anything but natural weighting
is used, it can have disastrous effects if a visibility is placed in a bin by
itself when most visibilities share their bin with several others.  Uniform
weighting would give the lone visibility as much weight as a large batch of
visibilities in a nearby bin, thereby amplifying the error of the lone
visibility \citep{bib:briggsphd}.  Smear fitting avoids binning by using
analytical Fourier transforms for its model components.  The final image is
displayed using pixels, but there is no error generated in the Fourier
transformation of the model.  The residuals still use the FFT, but typically in
the final image their dynamic range is so low that the introduced error is
negligable.

\subsection{Beam elongation}
\label{sec:beamelongation}

\begin{figure}
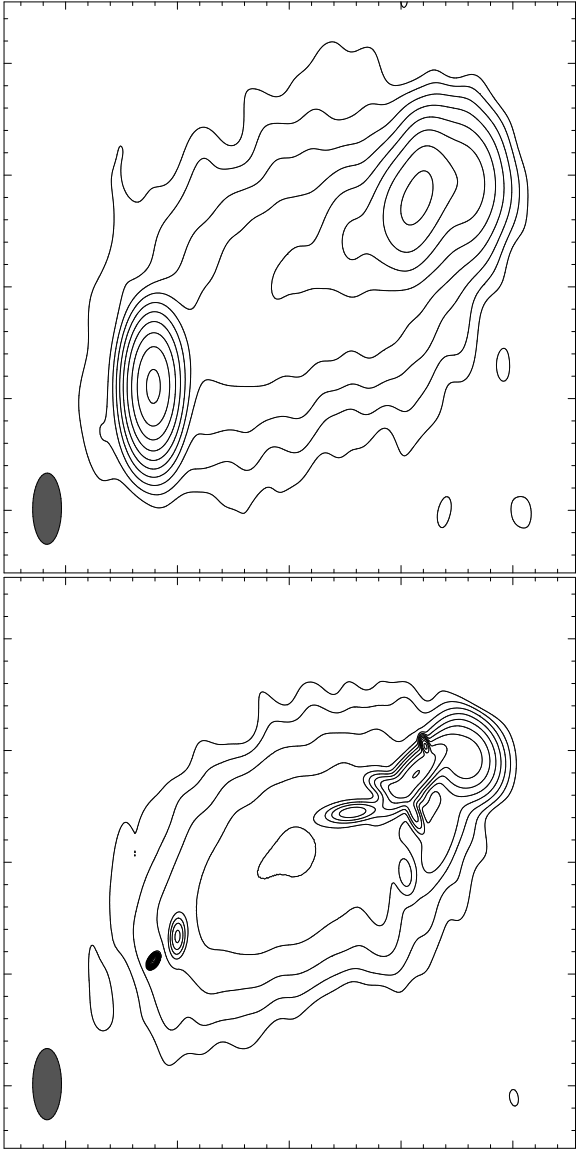

   \plotone{012435-0bsnap-30clnnat.ps}
   \plotone{012435-0bsnap-30sm.ps}
   \caption{Deconvolved images of a simulated VLA B array snapshot of
   Figure~\ref{fig:012435ref} at a declination of $-30^\circ$.  Top: CLEAN\@.
   Bottom: smear fitted.  The solid gray ellipses are the half maximum extent
   of the CLEAN beam (same in both) and the contours start at 25
   $\mu$Jy/arcsec$^2$, with each separated by a factor of 2 in brightness.}
   \label{fig:012435-0bsnap-30}
\end{figure}

A notable advantage of smear fitting is that it is less affected by the source
elevation angle than CLEAN (Figure~\ref{fig:012435-0bsnap-30}).  The axial
ratio of an east--west interferometer's dirty beam is the absolute sine of the
source's declination, meaning that the major axis approaches infinity for
objects near the celestial equator.  Similarly, an array with both east--west
and north--south baselines has foreshortened $uv$ coverage for low elevation
sources, so many observations have strongly elliptical dirty beams that distort
the appearance of their features.  CLEAN can mask that distortion by using a
round restoring beam with the same radius as the semimajor axis of the dirty
beam, but at the cost of losing resolution along the minor axis.  Smear fitting
copes better, since the density of baselines goes up when they are compressed
along one axis.  Thus a component that is resolved along that axis will have
more visibilities brought in to where it needs them, so its signal to noise
ratio will improve and it will not be smeared as much (see
Table~\ref{tab:sbdepondec}).  Less resolved features benefit less, since for
them the effect on \csq\ (i.e.\ resistance to smearing) of a visibility is
proportional to the fourth power of the baseline length, so the outer baselines
dominate (Appendix~\ref{sec:baselinelev}).  If the unresolved features are
bright their major axes will be strongly affected by the baseline
foreshortening, but if they are faint enough for the Fourier transform of their
smearing beams to be enclosed by the envelope of where there are visibilities,
they also experience the rounding effect.

\newcommand{\header}[1]{\multicolumn{1}{c|}{#1}}
\newcommand{\lheader}[1]{\multicolumn{1}{c}{#1}}
\newcolumntype{x}{D{.}{.}{4}}
\newcolumntype{y}{D{.}{.}{5}}
\begin{table}
  \centering
  \begin{tabular}{r|xx|yy}
      & \multicolumn{2}{c|}{Feature} & \multicolumn{2}{c}{Smearing beam}\\
    \header{$\delta$ (\degr)} & \lheader{FWHM (\arcsec)} &
    \header{$I\textsub{peak}/I\textsub{rms}$} & \lheader{FWHM (\arcsec)} &
      \lheader{Axial ratio}\\\hline
      &                             & 423.4   & 0.067 & 0.480 \\
      & \raisebox{1.5ex}[0pt]{0.20} &  43.23  & 0.235 & 0.523 \\\cline{2-5}
      &                             &  39.00  & 0.416 & 0.970 \\
      & \raisebox{1.5ex}[0pt]{2.00} &   3.90  & 1.57  & 0.949 \\\cline{2-5}
   \raisebox{5.8ex}[0pt]{-30} & \multicolumn{2}{r|}{CLEAN beam} &
                                                0.964 & 0.404 \\\hline
      &                             & 384.2  & 0.039 & 0.829 \\
      & \raisebox{1.5ex}[0pt]{0.20} &  38.36 & 0.131 & 0.850 \\\cline{2-5}
      &                             &  46.16 & 0.431 & 0.903 \\
      & \raisebox{1.5ex}[0pt]{2.00} &   4.08 & 1.21  & 0.961 \\\cline{2-5}
   \raisebox{5.8ex}[0pt]{0} & \multicolumn{2}{r|}{CLEAN beam}
                                              & 0.509 & 0.766 \\\hline
      &                             &  373.5 & 0.036 & 0.962 \\
      & \raisebox{1.5ex}[0pt]{0.20} &   37.09 & 0.126 & 0.975 \\\cline{2-5}
      &                             &   40.88 & 0.454 & 0.863 \\
      & \raisebox{1.5ex}[0pt]{2.00} &    4.13 & 1.367 & 0.921 \\\cline{2-5}
   \raisebox{5.8ex}[0pt]{30} & \multicolumn{2}{r|}{CLEAN beam}
                                              & 0.423 & 0.922 \\
  \end{tabular}
  \caption{Smearing beam dependence on source declination for simulated VLA A
    Array snapshots on the meridian.  Note that $I\textsub{peak}$ is the
    surface brightness in the \emph{dirty} map.}
  \label{tab:sbdepondec}
\end{table}

\subsection{Customized component types}
\label{sec:custom}

Elliptical gaussians are convenient basis functions for modelling most sources,
but some objects, especially ones that have steep edges and known physical
forms, are better fit if different functions are introduced.
Figure~\ref{fig:vy22} shows four deconvolutions of a VLA snapshot of a
planetary nebula, Vy2-2 \citep{bib:hces}.  Vy2-2 is young and well described
at 15 GHz by an optically thin shell.  The shell was only moderately resolved
by CLEAN, and smear fitting with elliptical gaussians either broke the symmetry
of the source or produced a unphysically negative center for the shell.  The
ring could be better approximated by using more gaussians, but splitting the
the flux into smaller portions would enlarge the smearing beam for each
component.  Ideally smear fitting should use as few parameters as necessary,
and the right framework (or model type) for those parameters.  The difmap
and the smerf patch do not explicitly have optically thin shells as a
component type (although optically thick ones are available as flat disks), but
one can be easily constructed by placing a negative optically thin ellipsoid
(OTE) inside a positive one.  Since the pair correspond to a single feature,
and especially since it is the difference of their flux densities, not their
individual flux densities, that is physically meaningful, smearing gaussians
were calculated for them simultaneously instead of serially.  The result was a
much better fit, and the possibility of measuring the geometrical thickness of
the shell.

\subsection{Future possibilities}
\label{sec:future}

Model fitting by minimizing \csq\ is extremely flexible, and could be extended
beyond providing a choice of component types.  One avenue for future work would
be to combine data from multiple polarizations and/or frequencies while
applying constraints (for example $Q^2 + U^2 + V^2\leq I^2$ and/or a spectral
index) on how the components should appear in each subset of the data.
Although such dynamic constraints have not yet been used with smear fitting, it
is already possible to construct a model using one set of data and use it as
the initial model with related data sets.  For example, the positions of a set
of isolated unresolved objects are the same for all polarizations and
frequencies, so a model of the set can be produced and smeared in Stokes I, and
then the Q and U models can be obtained by simply refitting the fluxes without
needing to change the smears.

\section{Conclusion}
\label{sec:conclusion}

Smear fitting is an image deconvolution method that fits a near maximally
simple model to interferometer visibilities and then broadens the model to
account for the uncertainty of those visibilities.  The model construction
method avoids several problems that can limit the quality of CLEAN and maximum
entropy deconvolutions, and smearing generally yields sharper and fairer images
than CLEAN.

As mentioned above, smear fitting has been implemented as a modification
(patch) to difmap, a well known program for imaging and selfcalibrating data
from radio interferometers.  The patch, known as smerf, is freely available
under the GNU license \citep{bib:gnu} at
http:$/\!$/www.drao-ofr.hia-iha.nrc-cnrc.gc.ca/$\!\sim$rreid/smerf/, and
includes a manual on its use.

\section*{Acknowledgments}

I thank Dr P. Kronberg for his helpful comments and support from NSERC grant
\#5713, and Martin Shepherd for making difmap's source freely available.  I am
grateful to Prof.\ E. Seaquist for kindly providing the Vy2-2 data, and Dr S.
Dougherty for offering many useful suggestions as an early user of the software
and reader of this paper.  I thank the anonymous referee for a thorough review
and useful comments.  Drs A. Gray and T. Landecker also provided helpful
comments on a draft of this paper.

\bibliographystyle{mn2e}
\bibliography{thesis}

\begin{thebibliography}{}

\bibitem[\protect\citeauthoryear{Bhatnagar \& Cornwell}{Bhatnagar \&
  Cornwell}{2004}]{bhatnagar04:aspen1}
Bhatnagar S.,  Cornwell T.~J.,  2004, Astronomy \& Astrophysics, 426, 747

\bibitem[\protect\citeauthoryear{Born \& Wolf}{Born \&
  Wolf}{1999}]{bib:bornwolf7th}
Born M.,  Wolf E.,  1999, Principles of Optics, 7 edn.
Cambridge University Press, pp 370--372

\bibitem[\protect\citeauthoryear{Briggs}{Briggs}{1995}]{bib:briggsphd}
Briggs D.~S.,  1995, PhD thesis, New Mexico Institute of Mining and Technology

\bibitem[\protect\citeauthoryear{{Christianto} \& {Seaquist}}{{Christianto} \&
  {Seaquist}}{1998}]{bib:hces}
{Christianto} H.,  {Seaquist} E.~R.,  1998, AJ, 115, 2466

\bibitem[\protect\citeauthoryear{Cornwell \& Holdaway}{Cornwell \&
  Holdaway}{2006}]{bib:multiscaleclean}
Cornwell T.,  Holdaway M.,  2006, submitted to Astronomy \& Astrophysics

\bibitem[\protect\citeauthoryear{Cornwell, Braun \& Briggs}{Cornwell
  et~al.}{1999}]{bib:cbb}
Cornwell T.~J.,  Braun R.,    Briggs D.~S.,  1999, in Taylor G.~B.,  Carilli
  C.~L.,   Perley R.~A.,  eds, Synthesis Imaging in Radio Astronomy {II}
  Vol.~180 of ASP Conference Series, {Deconvolution}.
pp 151--170

\bibitem[\protect\citeauthoryear{{Cornwell} \& {Evans}}{{Cornwell} \&
  {Evans}}{1985}]{CEMEM}
{Cornwell} T.~J.,  {Evans} K.~F.,  1985, Astronomy \& Astrophysics, 143, 77

\bibitem[\protect\citeauthoryear{{Cornwell} \& {Wilkinson}}{{Cornwell} \&
  {Wilkinson}}{1981}]{selfcal1981}
{Cornwell} T.~J.,  {Wilkinson} P.~N.,  1981, MNRAS, 196, 1067

\bibitem[\protect\citeauthoryear{{Gull} \& {Skilling}}{{Gull} \&
  {Skilling}}{1983}]{bib:gullskilling}
{Gull} S.~F.,  {Skilling} J.,  1983, in Indirect Imaging. Measurement and
  Processing for Indirect Imaging. Proceedings of an International Symposium
  held in Sydney, Australia, August 30-September 2, 1983. Editor, J.A. Roberts;
  Publisher, Cambridge University Press, Cambridge, England, New York, NY,
  1984. LC \# QB51.3.E43 I53 1984. ISBN \# 0-521-26282-8. P.267, 1983 {The
  Maximum Entropy Method}.
pp 267--+

\bibitem[\protect\citeauthoryear{{H{\" o}gbom}}{{H{\"
  o}gbom}}{1974}]{bib:hogbom74}
{H{\" o}gbom} J.~A.,  1974, \aaps, 15, 417

\bibitem[\protect\citeauthoryear{{H{\" o}gbom}}{{H{\"
  o}gbom}}{2003}]{2003ASPC..300...17H}
{H{\" o}gbom} J.~A.,  2003, in ASP Conf. Ser. 300: Radio Astronomy at the
  Fringe {Early Work in Imaging}.
pp 17--20

\bibitem[\protect\citeauthoryear{Narayan \& Nityananda}{Narayan \&
  Nityananda}{1984}]{bib:nn1984}
Narayan R.,  Nityananda R.,  1984, in Roberts J.~A.,  ed., Indirect Imaging
  Maximum entropy -- flexibility versus fundamentalism.
Cambridge University Press, Cambridge, England, pp 281--290

\bibitem[\protect\citeauthoryear{Pearson}{Pearson}{1999}]{bib:pearsonnida}
Pearson T.~J.,  1999, in Taylor G.~B.,  Carilli C.~L.,   Perley R.~A.,  eds,
  Synthesis Imaging in Radio Astronomy {II} Vol.~180 of ASP Conference Series,
  Non--imaging data analysis.
pp 335--354

\bibitem[\protect\citeauthoryear{Perley}{Perley}{1999}]{bib:hidrim}
Perley R.~A.,  1999, in Taylor G.~B.,  Carilli C.~L.,   Perley R.~A.,  eds,
  Synthesis Imaging in Radio Astronomy {II} Vol.~180 of ASP Conference Series,
  High dynamic range imaging.
pp 275--299

\bibitem[\protect\citeauthoryear{{Puetter} \& {Yahil}}{{Puetter} \&
  {Yahil}}{1999}]{bib:pixon99}
{Puetter} R.~C.,  {Yahil} A.,  1999, in ASP Conf. Ser. 172: Astronomical Data
  Analysis Software and Systems VIII {The Pixon Method of Image
  Reconstruction}.
pp 307--+

\bibitem[\protect\citeauthoryear{Reid}{Reid}{2003}]{bib:mythesis}
Reid R.~I.,  2003, PhD thesis, University of Toronto

\bibitem[\protect\citeauthoryear{{Reid}, {Kronberg} \& {Perley}}{{Reid}
  et~al.}{1999}]{1999ApJS..124..285R}
{Reid} R.~I.,  {Kronberg} P.~P.,    {Perley} R.~A.,  1999, ApJS, 124, 285

\bibitem[\protect\citeauthoryear{{Sault}, {Teuben} \& {Wright}}{{Sault}
  et~al.}{1995}]{bib:miriad1995}
{Sault} R.~J.,  {Teuben} P.~J.,    {Wright} M.~C.~H.,  1995, in ASP Conf. Ser.
  77: Astronomical Data Analysis Software and Systems IV {A Retrospective View
  of MIRIAD}.
pp 433--+

\bibitem[\protect\citeauthoryear{{Shepherd}}{{Shepherd}}{1997}]{bib:difmap1997}
{Shepherd} M.~C.,  1997, in A.S.P. Conf.\ Ser.\ 125: Astronomical Data Analysis
  Software and Systems VI Vol.~6, {Difmap: an Interactive Program for Synthesis
  Imaging}.
pp~77+

\bibitem[\protect\citeauthoryear{Stallman}{Stallman}{1991}]{bib:gnu}
Stallman R.~M., , 1991, GNU General Public License

\bibitem[\protect\citeauthoryear{Taylor, Gibson, Peracaula, Martin, Landecker,
  Brunt, Dewdney, Dougherty, Gray, Higgs, Kerton, Knee, Kothes, Purton,
  Uyan\i{}ker, Wallace, Willis \& Durand}{Taylor et~al.}{2003}]{bib:cgps2003}
Taylor A.~R.,  Gibson S.~J.,  Peracaula M.,  Martin P.~G.,  Landecker T.~L.,
  Brunt C.~M.,  Dewdney P.~E.,  Dougherty S.~M.,  Gray A.~D.,  Higgs L.~A.,
  Kerton C.~R.,  Knee L. B.~G.,  Kothes R.,  Purton C.~R.,  Uyan\i{}ker B.,
  Wallace B.~J.,  Willis A.~G.,    Durand D.,  2003, AJ, 125, 1350

\bibitem[\protect\citeauthoryear{{Wakker} \& {Schwarz}}{{Wakker} \&
  {Schwarz}}{1991}]{mrc}
{Wakker} B.~P.,  {Schwarz} U.~J.,  1991, in ASP Conf. Ser. 19: IAU Colloq. 131:
  Radio Interferometry. Theory, Techniques, and Applications {The
  multi-resolution clean}.
pp 268--271

\end{thebibliography}

\appendix

\section{Baseline leverage}
\label{sec:baselinelev}
Consider a feature modeled by a circular gaussian with flux $f$ and FWHM $a$.
We can without loss of generality place the phase tracking center at its
location, so that the model visibilities $V_{m,i}$ are $f\exp\left( -0.36
  a^2 u_i^2 \right)$, where $u_i$ is the length of baseline $i$.  Writing the
differences between the measured and model visibilities as ${\sf \epsilon}_i$,
\begin{equation}
  \label{eq:csqepsiloni}
  \chi^2\textsub{unsmeared} = \sum_i \left| {\sf \epsilon}_i / \sigma_i \right|^2
\end{equation}

Smearing the component with a round gaussian with FWHM $a_s$ multiplies the
model visibilities by $\beta(a_s^2 u_i^2) \equiv \exp\left( -0.36 a_s^2 u_i^2
\right)$.  (This discussion can easily be extended to an elliptical smearing
beams, but it is not warranted here.)  The expected rise in \csq\ due to
smearing is 
\begin{eqnarray}
  \Delta\csq & \equiv & \chi^2\textsub{smeared} - \chi^2\textsub{unsmeared}\\
             & = & \sum_i \frac{1}{\sigma_i^2}(1 - \beta)V_{m,i} \left[
              (1 - \beta)V_{m,i} + 2{\sf \epsilon}_i \right]
\end{eqnarray}

The leading term in a Maclaurin expansion of $1 - \beta$ is $0.36 a_s^2 u_i^2$,
so if the component is sharp (nearly flat in the $uv$ plane), the ${\sf
  \epsilon}_i$ are small, and there is little smearing, the effect of a
visibility on $\Delta\csq$ is proportional to the fourth power of its baseline
length.  If the component is resolved $V_{m,i}$ attenuates the importance of
the outer baselines.  Their special status can also be removed by a large
smearing beam, since $1 - \beta$ saturates at 1. 

Note that this leverage is simply the relative importance of the visibility.
Since $\beta$ is a function of the product of $a_s$ and $u_i$, $a_s$ is
inversely proportional to an average baseline length, as one would expect, but
the average is weighted towards the outer regions of where there is significant
measured amplitude in the $uv$ plane.

\label{lastpage}
\end{document}